\documentclass{desyproc}

\usepackage{feynmp}
\usepackage[usenames,dvipsnames]{xcolor}
\usepackage{slashed}
\usepackage[squaren]{SIunits}
\usepackage{feynmp}
\usepackage{hyperref}
\usepackage{cite}

\DeclareSymbolFont{rsfs}{U}{rsfs}{m}{n}
\DeclareSymbolFontAlphabet{\mathrsfs}{rsfs}

\newcommand{\lambdabar}{{\mkern0.75mu\mathchar '26\mkern -9.75mu\lambda}}
\newcommand{\plp}{\Delta\phi}

\begin{document}
\title{Volkov States and Non-linear Compton Scattering in Short and Intense Laser Pulses}

\author{{\slshape Daniel Seipt}\\[1ex]
Lancaster University, Physics Department, Lancaster LA1 4YB, United Kingdom \& \\
The Cockcroft Institute Daresbury Laboratory, Warrington WA4 4AD, United Kingdom }


\acronym{SFHQ2016} 
\doi  

\maketitle

\begin{abstract}
The collision of ultra-relativistic electron beams with intense short laser pulses makes possible to study QED in
the high-intensity regime.
Present day high-intensity lasers mostly operate with short pulse durations of several tens of femtoseconds, i.e.~only a few optical cycles.
A profound theoretical understanding of short pulse effects is important not only for studying fundamental aspects of high-intensity laser matter interaction, but also for applications as novel X- and gamma-ray radiation sources.
In this article we give a brief overview of the theory of high-intensity QED with focus on effects due to the short pulse duration.
The non-linear spectral broadening in non-linear Compton scattering due to the short pulse duration and
its compensation is discussed.
\end{abstract}

\section{Introduction \& High-Intensity Parameters}

Rapid advances in tera- and petawatt-class laser technology allow to explore light-matter interactions in the
experimentally uncharted high-intensity regime. 
Theoretical predictions include, for instance,
intensity dependent non-linearities in high-energy photon emission and pair production processes
\cite{Ritus:JSLR1985},
or non-linear quantum vacuum optics effects such as vacuum birefringence and photon splitting \cite{Dunne:IntJModPhys2012,Heinzl:OptCommun2006,Karbstein:PRD2015,Gies:PRD2016}.
Eventually one expects the spontaneous creation of particle anti-particle pairs from the vacuum (so-called Sauter-Schwinger effect)
for field strengths on the order of\footnote{We use natural Heaviside-Lorentz units with $\hbar=c=1$ and fine structure constant $\alpha = e^2/4\pi\simeq1/137$. Scalar products between four-vectors are denoted as
$(kp) = k\cdot p = k^\mu p_\mu = k^0p^0 - \mathbf k\cdot \mathbf p$.}
 $E_s\sim m^2/e\sim 1.3\times 10^{18} \,\rm V/m$ \cite{Schwinger:PR1951,Popov:PLA2002,Blaschke:PRL2006,Hebenstreit:PRD2008}.

When an electron interacts with a high-intensity laser, the laser pulse can be described as a plane wave.
Plane waves are null-fields, i.e. both field invariants vanish: $E^2-B^2 = \mathbf E\cdot \mathbf B =0$, which means no pairs can be produced from vacuum via the Sauter-Schwinger mechanism).
The majority of modern high-intensity laser systems produces laser pulses with wavelengths on the order of $1\,\micro\metre$,
and with pulse durations on the order of tens of femtoseconds, i.e. only a few optical cycles.
We therefore shall investigate in this article the important effects of the short pulse duration in high-intensity laser matter interactions.

Let us first recall some of the properties of (transverse) plane waves, which we assume to propagate along the negative $z$ axis,
and with frequency $\omega = \mathcal O(1\,\rm eV)$.
Their wave vector is a light-like four-vector, $k^2 = \omega^2-\mathbf k^2 = 0$, and
the field strength is a univariate function, $F^{\mu\nu} = F^{\mu\nu}(k\cdot x)$.
In this case, $\phi = k\cdot x = \omega x^+$, which tells us that the field depends solely on the light-front variable $x^+ = t+z$ (cf.~Appendix~A).
In addition, Maxwell’s equations in vacuum require that $F^{\mu\nu}$ is transverse, $k_\mu F^{\mu\nu} = 0$.
The field can be represented by a transverse vector potential, $A^\mu = (0, \mathbf A_\perp ( k\cdot x) , 0)$, with $k\cdot A=0$,
and parametrized as $\mathbf A_\perp = A_{0} g(\phi/\plp) {\rm Re} [\boldsymbol \epsilon e^{-i\phi}] $
with a complex polarization vector $\boldsymbol \varepsilon$, and an envelope function $g$ with pulse duration $\plp$.
The interaction of a probe electron with four-momentum $p^\mu$ with the plane wave field can be characterized by the following
dimensionless gauge and Lorentz invariant parameters:
(i) the classical nonlinearity parameter $a_0 = eA_0/m$ \cite{Heinzl:OptCommun2009}, 
(ii) the quantum energy parameter $b_0 = (kp)/m^2$, and
(iii) the quantum efficiency parameter $\chi_e = e \sqrt{(F\cdot p)^2} /{ m^3} = a_0 \, b_0$.
For a pulsed laser field one has in addition the dephasing parameter $a_0^2 \plp $
\cite{Hartemann:PRL2010}, which represents the ratio of the
laser peak intensity and the laser bandwidth $a_0^2 \plp \propto I /(\Delta \omega/\omega)$.

The classical nonlinearity parameter $a_0$ represents the laser energy density seen by the probe electron and
can be related to the laser intensity via
$a_0^2= 7.309\times 10^{-19} I [\watt\per\centi\metre^2] \lambda^2[\micro\metre]$.
With present laser technology one is able to reach light intensities on the order of
$I = \unit{2\times 10^{22} }{\watt\per\centi\metre^2} $,
which corresponds to $a_0\sim 100$ at $\lambda = \unit{800}{\nano\metre}$ typical for Ti:Sa laser systems \cite{Danson:HPLSE2015,Yanovsky:OptEx2008}.
With the upcoming generation of multi-petawatt lasers one expects to reach intensities up to
$ 10^{23}\ldots 10^{25}$ $\watt\per\centi\metre^2$ and even beyond
\footnote{For instance, the Vulcan laser at RAL-CLF (UK)
(\url{{http://www.clf.stfc.ac.uk/CLF/}}), the ELI project (Czech Republic, Hungary, Romania)
\url{{http://www.eli-beams.eu}}, or the OMEGA EP-OPAL project (Rochester, USA).}.
The parameter also $a_0$ represents the work done by the field in one wavelength $\lambdabar$ of the wave in units of the electron mass $a_0 = \frac{e E_0 \lambdabar }{m}$.
The limit $a_0\to \infty$, refers to both the limit of infinite intensity at fixed frequency and the static limit $\omega\to 0$ ($\lambda\to \infty$) at fixed intensity. For $a_0\gg 1$ the probabilities of processes happening in the laser field approach those in a constant crossed field \cite{Ritus:JSLR1985} and are described by the quantum efficiency parameter $\chi_e$ alone.
The regime $a_0 \gg 1$ is also referred to as ``quasi-static'' or ``tunnelling'' regime, i.e.~$1/a_0$ serves as
the Keldysh parameter of the electron laser interaction.

The quantum efficiency parameter $\chi_e$ is the value of the electric field experienced
by the particle in its rest frame in units of the Sauter-Schwinger field strength, $\chi_e = 2\gamma E/ E_s$, with $E_s = m^2/e$.
This means that $\chi_e$ can be large for ultra-relativistic particles with $\gamma\gg1$ colliding with a laser pulse, although the electric field in the laboratory frame is much weaker than $E_s$ due to the Lorentz transformation of the transverse electric field.

\section{Classical Dynamics of an Electron in a Laser Pulse}

\label{sect.classical}

{Before we dive into the quantum theory of the laser matter interaction let us start with
a classical description. The classical picture is valid whenever both the quantum energy parameter $b_0\ll1$ and the quantum efficiency parameter $\chi_e\ll1$.}
According to the classical theory of electrodynamics,
the dynamics of point-like charged particles in a given external field configuration, described by the field strength tensor $F_{\mu\nu}= \partial_\mu A_\nu - \partial_\nu A_\mu$, is governed by the Lorentz force equation \cite{book:Jackson}
\begin{align}
m \, \frac{d u^\mu}{d\tau} & =  F^{\mu \nu}(x) u_\nu\,, \label{eq:lorentz:force}
\end{align}
where $m$ is the electron mass and we have conveniently absorbed the electron charge into the background field vector potential $eA \to A$.
The electron's four-velocity $u^\mu =  d x^\mu/d\tau$ is the tangent vector to the particle's world line $x^\mu(\tau)$, parametrized by
proper time $\tau $. Due to the anti-symmetry of the field strength tensor the four-acceleration $d u^\mu/d\tau$ is always orthogonal to the four-velocity, and the four-velocity is normalized as $u \cdot u= 1$.

In general this is a non-linear differential equation as the field strength has to be taken along the world line of the particle 
to be solved for $F^{\mu \nu} = F^{\mu\nu} ( x(\tau) )$.
The solution of \eqref{eq:lorentz:force} can be quite a difficult task.
For instance, the motion of an electron in standing waves can show chaotic behaviour \cite{Bauer:PRL1995}.
Luckily, the dynamics of an electron in a plane wave laser field is one of the few exactly solvable cases.

\subsection{Solution of the Lorentz Force Equation in Plane Waves}

The solution of the Lorenz force equation in a plane wave field becomes particularly simple when employing light-front coordinates (cf.~Appendix~A),
as one can find a very simple relation between the laser phase $\phi = \omega x^+$ and the particle's
proper time $\tau$.
By dotting the laser wave-vector into the equation of motion \eqref{eq:lorentz:force}
we immediately find $\frac{d}{d\tau} (k\cdot u) = 0$.
That means $k\cdot u = const.$ is a constant of motion, i.e.~the $u^+$ component of the electron's four-velocity is conserved
$u^+ = u^+_0$.
From this we find that the relation between the laser phase $\phi$, light-front time $x^+$ and proper time $\tau$ can be expressed as
\begin{align}
\label{eq:propertime:phase}
\frac{d\phi}{d\tau} &= k\cdot u = \omega u^+  \quad \Rightarrow \quad x^+ =  u^+ \, \tau \,.
\end{align}
That means one can replace the proper time derivative by a light-front time derivative in the equations of motion.
Since the normalized laser vector potential has only transverse components, $a^\mu = e A^\mu/m= (0,0,\mathbf a_\perp(x^+))$,
the transverse components of the Lorentz force equation can be written as
\begin{align*}
\frac{d \mathbf u_\perp}{d\tau} = -  (k\cdot u) \, \frac{d \mathbf a_\perp}{d\phi} \,,
\end{align*}
which gives $\mathbf u_\perp + \mathbf a_\perp = const = \mathbf u_{\perp,0}$ because of \eqref{eq:propertime:phase}.
Because the four-velocity is normalized, $u^2=1$, the fourth component $u^-$ of the velocity can be obtained simply via
\begin{align*}
u^-(\phi) = \frac{1+\mathbf u_\perp^2(\phi) }{u^+} = \frac{1+ [\mathbf u_{\perp,0} - \mathbf a_\perp(\phi) ]^2 }{u^+} 
\end{align*}
The particle world line $x^\mu$ is obtain by integrating the four-velocity again
\begin{align}
x^\mu & = \int \! d\tau \: u^\mu  = \int \! \frac{dx^+}{u^+} \: u^\mu \,.
\end{align}

For later use we here also define the \emph{kinetic four momentum} of the electron as $\pi^\mu = m u^\mu$.
The solution for the kinetic momentum can also be represented in a covariant way
as
\begin{align} \label{eq:pi}
\pi^\mu(x^+) = \Lambda^\mu_{\ \nu}(x^+) \, p^\nu \,,
\end{align}
where $p^\nu$ is the initial value of the momentum, and $\Lambda^\mu_{\ \nu}(\phi) = (e^{X})^\mu_{\ \nu}$
with
$$
X^{\mu\nu} 
= \int^{x^+} \! \frac{d z^+}{p^+} \, F^{\mu\nu}(z^+)
= \frac{k^\mu A^\nu - k^\nu A^\mu}{k\cdot p}
$$
has the form of a Lorentz transformation \cite{Brown:PRL1983,Brown:PRD1984}.
In particular, the minus-component of the kinetic electron momentum can be written as
\begin{align} \label{eq:piminus}
\pi^-  = p^- + \frac{2 \,p\cdot A - A^2}{p^+} \,.
\end{align}

\subsection{Radiation Back-Reaction}
\label{sect:RR}

When the electron interacts with a high-intensity laser pulse the back-reaction of the emission of radiation on
the electron motion, the so-called radiation reaction (RR), can become significant.
The radiation power is given by the Larmor formula, $P = - \frac{2\alpha}{3} \, \dot u^2 $. 
Whenever the emitted radiation energy $\sim\alpha \gamma^2 \omega a_0^2 \plp$ becomes comparable
to the electron's initial energy $\gamma m$, the equations of motion \eqref{eq:lorentz:force}
have to be amended by a radiation force term $F_R^\mu$:
\begin{align} \label{eq:RR:force}
m \dot u^\mu = F^{\mu\nu} u_\nu  + F_R^\mu \,.
\end{align}
{In the radiation dominated regime, characterized by the parameter $R_C=\alpha \chi_e a_0 \gtrsim 1$,
the electron loses a large fraction of it's kinetic energy in one cycle of the background laser field, and the RR force $F_R^\mu$ becomes comparable to the Lorentz force.
This regime could be reached, for instance by colliding a 500 MeV electron beam with a laser pulse with $a_0=200$,
which corresponds to a peak intensity of $10^{23} \rm W/cm^2$.}

Many different forms for the RR force $F_R^\mu$ have been suggested since the seminal works of Lorenz and Abraham more than a century ago \cite{Abraham:AnnPhys1902,Dirac:ProcRoySoc1938,Sokolov:PhysPlas2009,Ilderton:PLB2013}.
A particularly popular form of the RR force (which is often used in numerical simulations of laser-matter interaction at high intensity \cite{Capdessus:PRE2012,Vranic:CPC2016,Tang2016})
is the so-called Landau-Lifshitz force \cite{book:Landau2}.
It reads
\begin{align} \label{eq:RR:force:LL}
F^\mu_{R,LL} &=  
 \frac{2\alpha}{3m}
\left[
(u^\alpha \partial_\alpha F^{\mu\nu}) u_\nu
+ \frac{1}{m} F^{\mu\nu} F_{\nu\alpha} u^\alpha
+ \frac{1}{m} (F^{\alpha\nu}u_\nu)^2 u^\mu
 \right] \,.
\end{align}
The equations of motion \eqref{eq:RR:force} with the LL radiation force \eqref{eq:RR:force:LL} also possess closed form
analytic solutions in the presence of plane wave backgrounds \cite{DiPiazza:LMP2008}.
{By dotting again the laser wave-vector $k$ into the equations of motion we find
\begin{align*}
 \frac{d (k\cdot u) }{d\tau} =  \frac{2\alpha}{3m}  (a'\cdot a') \, (k \cdot u)^3
\end{align*}
which can be integrated to find a (non-linear) relation between the laser phase $\phi$ and proper time $\tau$ [cf.~\eqref{eq:propertime:phase}] allowing to integrate \eqref{eq:RR:force} \cite{DiPiazza:LMP2008}.}%

Finally, we note that the classical RR is a continuous process, in contrast to quantum RR which is stochastic.
Quantum RR effects are important whenever already the emission of a single photon can alter the electron trajectory significantly
due to the momentum recoil \cite{DiPiazza:PRL2010,Neitz:PRL2013,Blackburn:PRL2014}.
The transition from the classical to the quantum RR regime is characterized by $\chi_e \gtrsim 1$ and $R_Q = \alpha a_0 \gtrsim 1$ \cite{DiPiazza:PRL2010}.

\section{High-intensity QED in the Furry Picture}

\label{sect:volkov}

In high-intensity QED the laser field is usually described as a background field\footnote{However, the development of avalanche type QED cascades at extreme intensities
could cause a depletion of the laser, turning an initially strong field weak, and rendering the background field
approximation invalid \cite{Fedotov:PRL2010,Bulanov:PRL2010,Zhang:NJP2015,Seipt:2016}. See also A.~M.~Fedotov's contribution in this volume.}.
This background field has to be treated to all orders, which can be achieved by going to the Furry interaction picture
in which the background field is treated as part of the unperturbed system \cite{Furry:PR1951}.
{
For this we need to know the
solutions of the Dirac equation
\begin{align}
[ i\slashed\partial - \slashed A(\phi) - m] \Psi(x) & =  0 \label{eq.dirac}
\end{align}
in a given background $A$ describing the laser field.
The solutions $\Psi(x)$ of \eqref{eq.dirac}
in a univariate null-field background $A=A(\phi)$ are called Volkov wave functions \cite{Volkov:ZPhys1935}.
We use here the Dirac slash notation $\slashed p = p^\mu \gamma_\mu$ to denote
scalar products between four-vectors and the Dirac matrices $\gamma_\mu$.

\subsection{Derivation of Volkov States}

Let us now explicitly calculate the Volkov wave functions.
Although the derivation can be found in textbooks (e.g.~in~\cite{book:Landau4})
will be present it here in some detail to work out why a null-field background is required
in order to solve Eq.~\eqref{eq.dirac} and indicate possible extensions to more general backgrounds.
In the first step on transforms \eqref{eq.dirac} into a second order differential equation by multiplying with the adjoint Dirac operator
$i\slashed \partial - \slashed A + m $ from the left \cite{book:Landau4}, yielding
\begin{align}
\left[ (i \partial - A)^2 - m^2  - \frac{1}{2} \sigma^{\mu\nu} F_{\mu\nu}(x) \right] \Psi (x) = 0 \,,
\label{eq:dirac:2}
\end{align}
with $\sigma^{\mu\nu} = \frac{i}{2} [\gamma^\mu , \gamma^\nu]$ as the commutator of gamma matrices.
The so-called {Pauli interaction}
term $\frac{1}{2} \sigma^{\mu\nu} F_{\mu\nu}(x)$ describes the interaction of the half-integer electron spin with
the background field. 
The general idea is to seek a solution of \eqref{eq:dirac:2} in the form
$\Psi(x) = e^{-i(px)} \Omega(\phi) u$ with a constant four-vector $p^\mu$,
and where $\Omega$ is a ($4\times 4$) Dirac matrix depending only on the laser phase $\phi = k\cdot x$
and $u$ denotes a Dirac bi-spinor.

When switching off the background field, the quantum numbers $p^\mu$ represent the four-momentum of a free particle.
With the background field present, they represent the asymptotic electron momentum outside the laser pulse.
Because three momentum operators
$p^1= - i {\partial}/{\partial x^1}$,
$p^2= - i {\partial}/{\partial x^2}$ and
$p^+= 2 i {\partial}/{\partial x^-}$ commute with the Dirac-operator, the corresponding quantum numbers are conserved.
In addition, $p^- = (\mathbf p_\perp^2 +m^2)/p^+$ is determined by the mass shell condition $p^2=m^2$.
Similarly, the Dirac bi-spinors are the the free Dirac spinors $u_{p,r}$ for (on-shell) momentum $p$, fulfilling
$(\slashed p-m) u_{p,r}=0$, and where $r=1,2$ is the spin quantum number.
We choose here and in the following the normalization $\bar{u}_{p,r}u_{p,r'}=2m\,\delta_{rr'}$.

By plugging the above ansatz for the wave function into \eqref{eq:dirac:2} one obtains an equation for the unknown matrix function $\Omega$.
Because Eq.~\eqref{eq:dirac:2} is a second order differential equation, in general the equation for $\Omega$
can be expected to be a second order as well. However, the coefficient of the second order term turns out to be $k^2$.
That means for a light-like univariate background field with wave vector $k^2=0$ the equation for $\Omega$ is of first order.
Note that this is not true for space-like ($k^2<0$) or time-like ($k^2>0$) wave vectors.
The first case, for instance, corresponds to magnetic undulators or the wave propagation in a medium with refractive index $n_r>1$.
The latter case may refer to time dependent electric fields, or the propagation of waves in a medium with refractive index $n_r<1$, e.g.~a plasma.
No general solutions of relativistic wave equations exist in background fields with $k^2\neq 0$.
Recent attempts to find (approximate) solutions can be found in Refs.~\cite{Heinzl:PRD2016,Raicher:PRA2013,Raicher:PLB2015,Mendonca:PRE2011}.
The situation becomes even more complicated for \textit{bi-variate} backgrounds, for instance
for the electron dynamics in two counter-propagating laser beams \cite{King:2016}.

For the Volkov problem, with $k^2=0$, we eventually find the following \textit{first order} ordinary differential equation for the hitherto unknown matrix function $\Omega$:
\begin{align}
2i(kp)\, \frac{d\Omega}{d\phi} &= \left( 2 (pA) - A^2 - i\slashed k \slashed A' \right) \Omega \,.
\end{align}
This equation can be easily integrated yielding
\begin{align}
\Omega_{p}(\phi) &= \left( 1 + \frac{ \slashed  k \slashed A}{2(kp)} \right)
\, e^{-i f_{p} (\phi) } \,, \\
f_{p} (\phi) &= \int \frac{d \phi}{2k\cdot p} \left[ 2 p\cdot A(\phi) - A^2(\phi) \right]
= - \frac{p^-x^+}{2} + \frac{1}{2} \int \! dx^+ \, \pi^-(x^+) \label{eq:fp}
\end{align}
where we used that $\slashed k \slashed A$ is a nilpotent operator of grade 2, i.e.
$(\slashed k \slashed A)^{n\geq2}=0$, which again holds because the
background field has a light-like wave vector with $\slashed k\slashed k=k^2=0$ and $(kA)=0$.
The non-linear phase $f_{p}$
can be written as an integrals of the minus component of the classical kinetic four-momentum \eqref{eq:piminus}
by using that $\phi = \omega x^+$.
Note that the Volkov wave functions \eqref{eq:volkov.state}
are normalized using a covariant light-front normalization,\footnote{
The scalar product
between any two spinor wave functions
is defined in a Lorentz invariant way as \cite{book:Bagrov}
$(\bar\Psi_1|\Psi_2) = \int_\sigma d\sigma_\mu \bar \Psi_1(x) \gamma^\mu \Psi_2(x)$, with
$\sigma$ being a hypersurface in Minkowski space and $d\sigma_\mu$ the infinitesimal normal vector thereupon.
The hypersurface can be expressed in general curvilinear coordinates $\xi^\mu(x)$, where $\xi^0=const.$
defines the hypersurface and $\xi^1,\xi^2\xi^3$ parametrize $\sigma$,
such that $d\sigma_\mu = \frac{d\xi^0(x)}{\partial x^\mu} \sqrt{-g} d\xi^1 d\xi^2 d\xi^3$ with the determinant $g$ of the metric tensor in the coordinates $\xi^\mu$. Choosing a light-front surface $x^+ = const.$ yields \eqref{eq:scp}.
}
\begin{align} \label{eq:scp}
(\bar \Psi_{p',r'} | \Psi_{p,r}) = \frac{1}{2} \int \! dx^- d\mathbf  x^\perp \: \bar \Psi_{p',r'}(x) \gamma^+ \Psi_{p,r}(x) 
= 2p^+ (2\pi)^3 \delta(\mathsf p' - \mathsf p) \delta_{r'r} \,
\end{align}
and where $\sf p$ stands for $(p^+,\mathbf p^\perp)$, i.e.~$\delta(\mathsf p) = \delta(p^+)\delta(p^1)\delta(p^2)$.

It is also convenient to define the so-called Ritus matrices 
$E_p(x) = e^{-i(px)} \Omega_{p}(\phi)$,
such that the Volkov states
can be written as 
\begin{align} \label{eq:volkov.state}
\Psi_{p,r}(x) = E_p(x) u_{p,r} \,.
\end{align}
The Ritus matrices can decomposed as
$E_p = V(\phi)  \, e^{-iS_\mathrm{HJ}}$,
where $S_\mathrm{HJ}$ denotes the classical Hamilton-Jacobi action of a classical particle in the background field.
That means Volkov wave function is an exact semiclassical wavefunction, i.e.~while being an exact solution of the wave equation \eqref{eq.dirac} the action does not contain quantum corrections (in powers of $\hbar$).
The matrix
$V(\phi)$ is the bi-spinorial representation of the Lorentz transformation $\Lambda^{\mu}_{\ \nu}(\phi)$
that generates the classical orbits~\eqref{eq:pi}, and which is defined by
$$  V^{-1}\, \gamma^\mu  V = \Lambda^{\mu}_{\ \nu} \gamma^\nu \,.$$
The Lorentz transformation $V(\phi)$ transports the spinor $u_{p,r}$ along the classical trajectory \cite{Brown:PRL1983,Brown:PRD1984}.
The Ritus matrices, analytically continued to off-shell momenta $p^2\neq m^2$, have the following orthogonality and completeness properties:
\begin{align*}
\int \! d^4x \: \bar E_{p'}(x) E_p(x) &= (2\pi)^4 \, \delta( p'-p) \,, \qquad & &
\int \! d^4p \: E_p(x') \bar E_p(x) = (2\pi)^4 \, \delta(x-x') \,.
\end{align*}

The wave functions (\ref{eq:volkov.state}) represent the positive energy solutions $\Psi_{p,r} \equiv \Psi_{p,r}^{(+)}$ of (\ref{eq.dirac}),
i.e.~they describe the propagation of electrons in a background field.
In order to describe positrons as well (for instance to calculate the probabilities for pair production in a laser field)
we need the negative energy solutions of \eqref{eq.dirac}.
They can be obtained from (\ref{eq:volkov.state})
via the transformation $p \to - p$, i.e.~$\Psi_{p}^{(-)}(x) \equiv \Psi_{-p}(x)$,
where the notation for negative energy spinors is as usual $u_{-p} \equiv v_p$ \cite{Nousch:PLB2012,book:Itzykson}.

\subsection{Properties of Volkov States}

The properties of the Volkov states can be studied by investigating the
Ritus matrix function $E_{p}(x)$ as they contain all the information on the interaction with the laser field.
Since the $E_p(x)$ are $4\times4$ matrices it is reasonable
to study the different projections onto the basis elements of the Clifford algebra \cite{book:Peskin}.
It turns out that only the scalar and anti-symmetric tensor projections yield non-zero results:
\begin{align}
\mathcal S[E_p] &= \frac{1}{4} {\rm tr} \, E_p(x) = \exp \{i S_p(x) \}\, , \\
\mathcal T^{\mu\nu} [E_p] &= \frac{1}{4} {\rm tr} \, \sigma^{\mu\nu} E_p(x)
= \frac{i }{2k\cdot p} (A^\mu k^\nu - A^\nu k^\mu) \exp \{ i S_p(x) \} \,, \label{eq:projection.tensor}
\end{align} 
where $\sigma^{\mu\nu} = \frac{i}{2} \, [\gamma^\mu,\gamma^\nu]$, and tr denotes the trace over the
Dirac matrix indices.

\begin{figure}[!t]
\begin{center}
\includegraphics[width=0.51\textwidth]{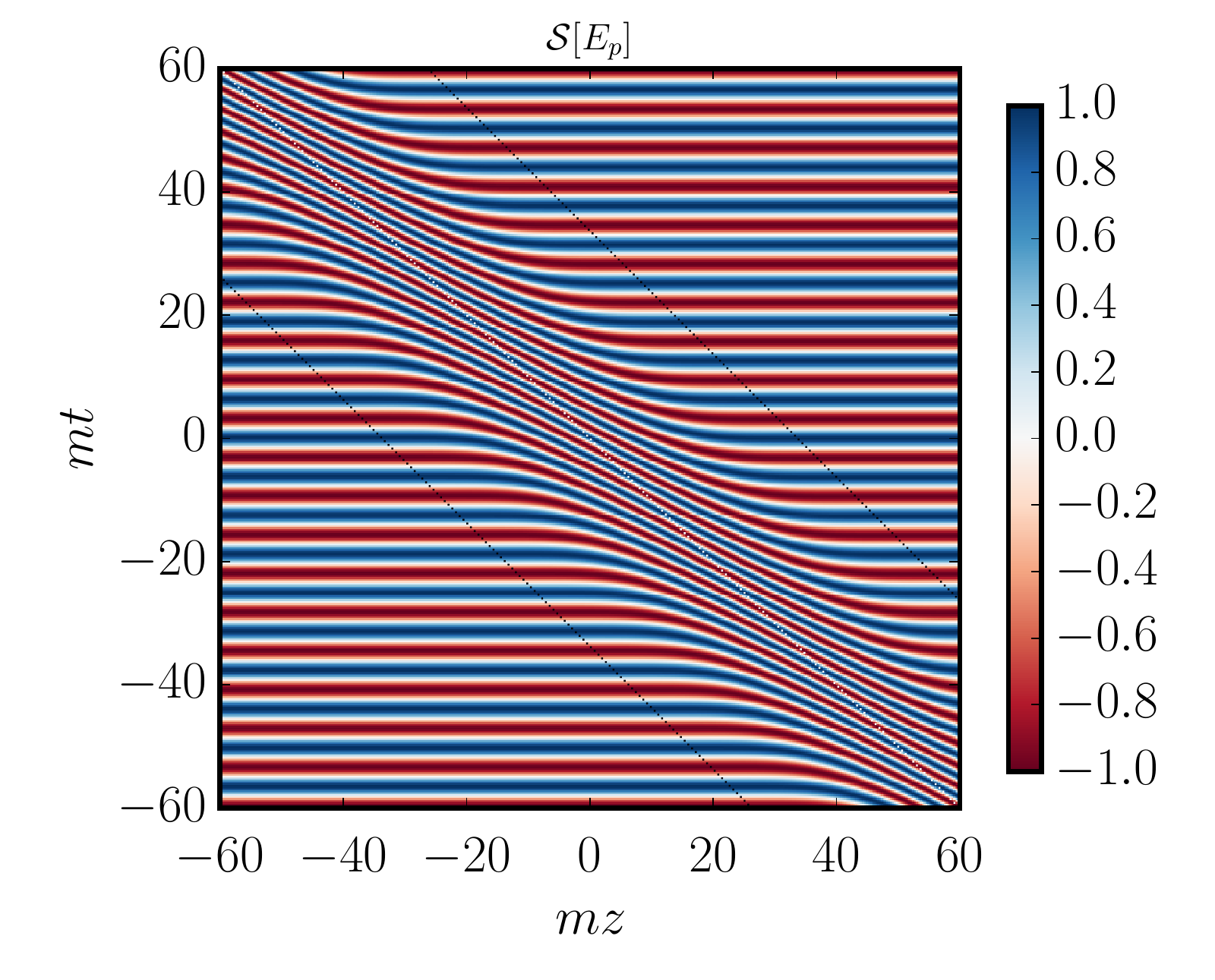}
\hspace*{-0.04\textwidth}
\includegraphics[width=0.51\textwidth]{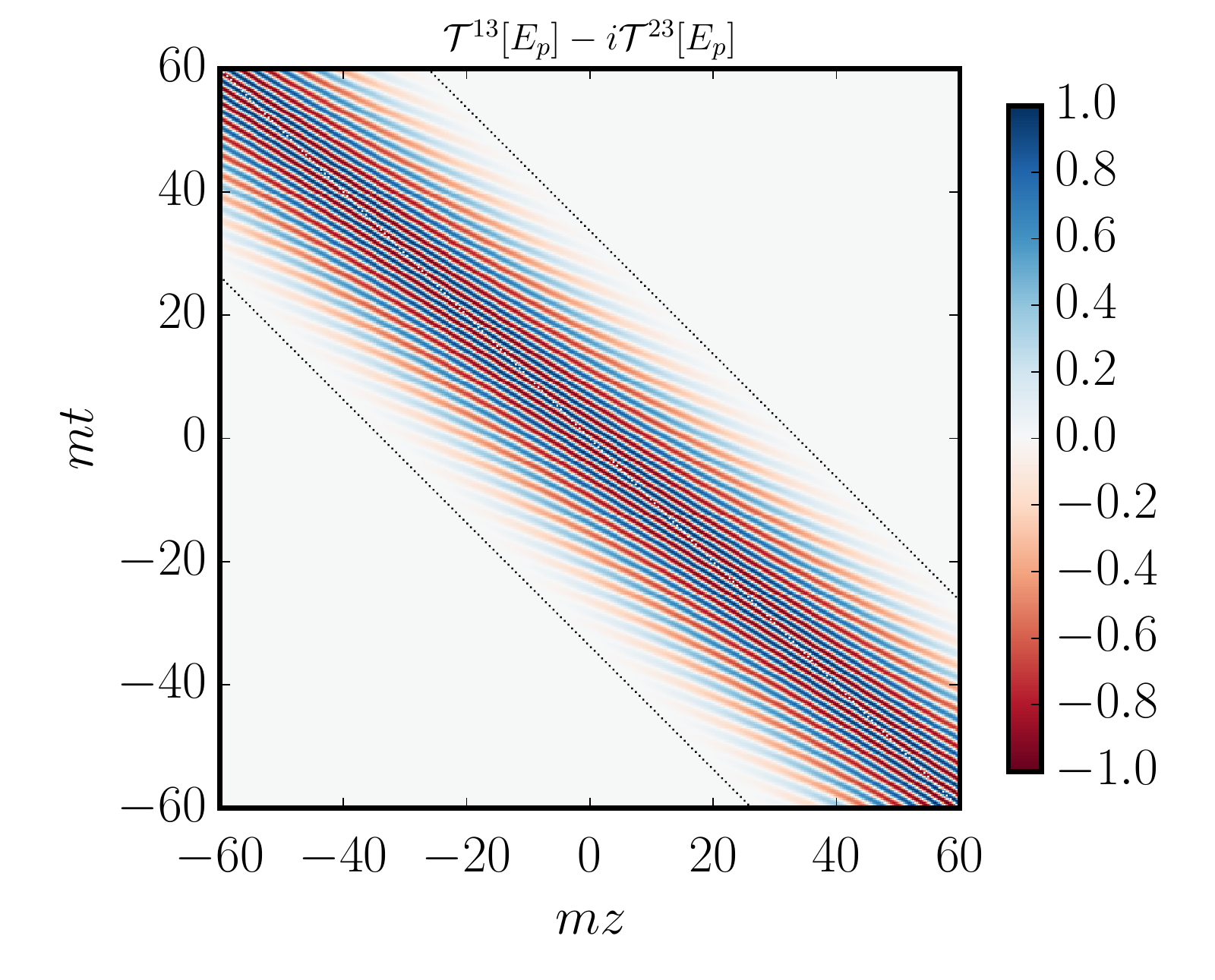}
\end{center}
\caption{
Contour plot of (the real part of) the scalar and tensor projections
of the Ritus matrices $E_p(x)$ in the ($z$-$t$) plane for a circularly polarized laser pulse with $a_0=2$
and pulse duration $\plp=20$.
}
\label{fig:volkov:state:spatial} 
\end{figure}

The real parts of the scalar projection $\mathcal{S} [E_p]$ and the combination of tensor projections
$\mathcal T^{13} -i \mathcal T^{23}$ are exhibited in Figure~\ref{fig:volkov:state:spatial}
for an electron with an energy of
$\unit{50}{\giga\electronvolt}$ propagating head-on through a strong laser pulse with
$a_0 = 2$ and a pulse duration $\plp=20$.
The two projections  are shown in the frame where the electron is initially at rest.
In that frame, the free electron wave function outside the laser pulse
behaves as $\propto \exp \{-i p\cdot x \} =\exp \{-i m t\}$.
In the case of the scalar projection of the Ritus matrices the
effect of the laser pulse is a local tilt of the electron wave fronts, see left panel of Fig.~\ref{fig:volkov:state:spatial}.
This behaviour corresponds to the build-up of an intensity dependent ponderomotive quasi-momentum inside the laser pulse \cite{Narozhnyi:JETP1996}.
The tensor projection of the Ritus matrices
shown in the right panel of Fig.~\ref{fig:volkov:state:spatial} are non-zero only in regions where the laser pulse is present.
They correspond to the $\slashed k \slashed A$ term in the Volkov states, i.e.~the Pauli interaction term.

In the following we construct wave packets from the Volkov wave functions and show that the motion of the centroid of the packet
follows is closely related to the classical trajectory \cite{Neville:PRD1971,Zakowicz:JMP2005}.
We will restrict the discussion to the scalar part of the Volkov wave function which describes charged spin-0 bosons.
As a simple example it suffices to construct a 
one-dimensional wave packet as a superposition of Volkov waves with a light-front momentum distribution $h_+(p^+)$ but all having the
same transverse momenta $\mathbf p_\perp=0$,
\begin{align}
\chi(x) = \int \! \frac{d p^+}{2p^+} \, h_+(p^+) \, {\mathcal S}[E_p(x)]\,.
\end{align}
Such a wave packet has a finite temporal duration, but remains infinitely extended in the transverse coordinates.
In the following we choose a Gaussian distribution
$h_+ = \mathcal N e^{- (p^+ - p^+_0)^2/ 2 \Delta^2}$ with Gaussian width $\Delta = 0.05\,m$
and with $p^+_0=m$.
The laser pulse parameters are the same as for Figures \ref{fig:volkov:state:spatial}.
This construction provides a localized wave packet in the ($t$-$z$) plane with a minimum Gaussian size
at $t=z=0$. This wavepacket is shown in Figure~{\ref{fig:volkov:wavepacket}} in the ($t$-$z$) plane.
The centroid of the wave packet follows the classical trajectory $z(t)$ (black curve),
which shows the close correspondence between the classical trajectory solutions and the Volkov wave functions.
In addition, the Dirac-Volkov current
\begin{align}
j^\mu &= \frac{ \bar \Psi_{p,r} \,\gamma^\mu \, \Psi_{p,r}}{  \bar \Psi_{p,r}   \Psi_{p,r} } = \frac{\pi^\mu(\phi)}{m}   = u^\mu(\phi)\,,
\end{align}
coincides with the classical four-velocity $u^\mu$.

\begin{figure}[!t]
\begin{center}
\includegraphics[width=0.52\textwidth]{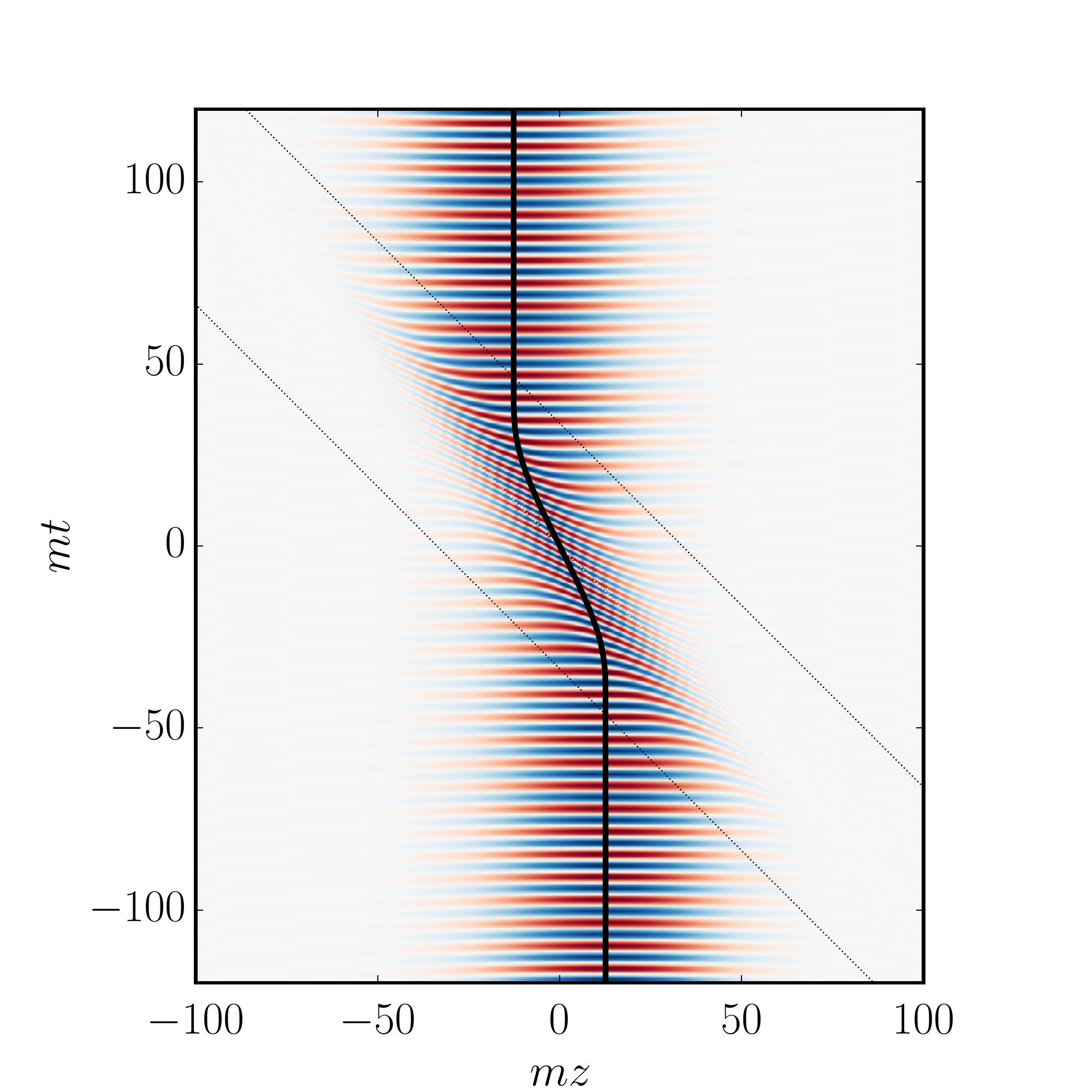}
\end{center}
\caption{Contour plot of a normalized scalar Volkov wave packet in
the ($t$-$z$) plane. For comparison, the classical trajectory is depicted as solid black curve.
The laser pulse propagates between the two black dotted lines.}
\label{fig:volkov:wavepacket}
\end{figure}

Let us now investigate the momentum space spectrum of the Volkov states and find relations to
the classical motion. The Ritus matrices $E_p(x)$ have a spectral representation as a superposition of plane waves
\begin{align}
 E_p(x) & =  \int \! \frac{d \ell}{2\pi} \, e^{-i( p +\ell k)\cdot x }\, \mathcal E_{p}(\ell)
\end{align}
with the spectral components
\begin{align}
\mathcal E_p(\ell) &=  \mathrsfs K_{0}(\ell)
 			+\frac{m a_0 \slashed k }{2( k p) } \big[\slashed\epsilon \, \mathrsfs K_{+}(\ell)
 										+\slashed\epsilon^*\, \mathrsfs K_{-}(\ell)\big] \,.
 										\label{eq:def.reduced.ritus}
\end{align}
which are be represented by the three scalar functions
\begin{align}
\left\{\begin{matrix}
\mathrsfs K_{0}(\ell) \\
\mathrsfs K_\pm(\ell)
\end{matrix}
\right\}
 & =\int d \phi 
 \left\{
 \begin{matrix}
 1 \\
 g(\phi) e^{\mp i( \phi + \hat \phi )}
 \end{matrix}
 \right\}  \exp \{ i \ell \phi - if_p(\phi) \}\,.
 \label{eq.def.Kn}
\end{align}
The variable $\ell$ parametrizes the amount of light-front laser momentum $k^-$ which is
exchanged between the electron and the background field.
It can be interpreted as some continuous analogue to the number of exchanged laser photons.

Let us now investigate the limit of infinitely long monochromatic plane wave background fields (IPW),
formally achieved by setting $g\to 1$.
In that case the background is a periodic function such that the Floquet theorem applies \cite{Floquet:1883}:
The solution of the Dirac equation \eqref{eq.dirac} in a periodic background takes the form
\begin{align}
\Psi_p(\phi) = e^{-i q\cdot x} \: \Phi(\phi)\,, \label{eq:floquet}
\end{align}
where $\Phi(\phi+2\pi) = \Phi(\phi)$ has the same periodicity as the background,
and $q^\mu = p^\mu + \frac{m^2 a_0^2}{4 k\cdot p} k^\mu$ is called the quasi-momentum.
The Fourier-zero mode, i.e.~the non-periodic part of the non-linear phase $f_p$ in \eqref{eq:fp},
has been absorbed into the definition of the quasi-momentum.

In the limit of an IPW background
the momentum distribution functions (\ref{eq.def.Kn}) degenerate to a delta comb
\begin{align}
\mathrsfs K_j(\ell) & 
						  \xrightarrow{g\to 1}
 			\sum_{n=-\infty}^\infty \delta(\ell - n - \frac{m^2a_0^2}{4k\cdot p} )
K_j(n)
\end{align}
with support at discrete momentum values $q^\mu + n k^\mu$.
For arbitrary polarization of the background field the coefficients $K_j(n)$ are related to generalized Bessel functions
(see e.g.~\cite{Korsch:JPA2006,Loetstedt:PRE2009}), and they turn into
ordinary Bessel functions of the first kind in the case of circular polarization.
%
The Volkov wave function in an IPW background appears as an infinite sum over discrete momentum states
\begin{align}
E_p(x) &= \sum_{n=-\infty}^\infty e^{-i(q + n k)\cdot x} \: \mathcal E_p^{\rm IPW}( n ) \,,
\label{eq.sum.zeldovich}
\end{align}
 which are also called Zel'dovich levels \cite{Zeldovich:JETP1967}.
{The level structure furnishes an easy interpretation of strong-field phenomena.
For instance, the appearance of harmonics in the non-linear Compton spectra
or the resonant singularities in second-order strong-field scattering processes can be seen as transitions between
different Zel'dovich of the incident and final state Volkov electrons.}
Modifications of this level structure due to radiative corrections, i.e.~the electron self-energy have been calculated as well \cite{Becker:JPA1976}.

\begin{figure}[!t]
\includegraphics[width=0.49\columnwidth]{{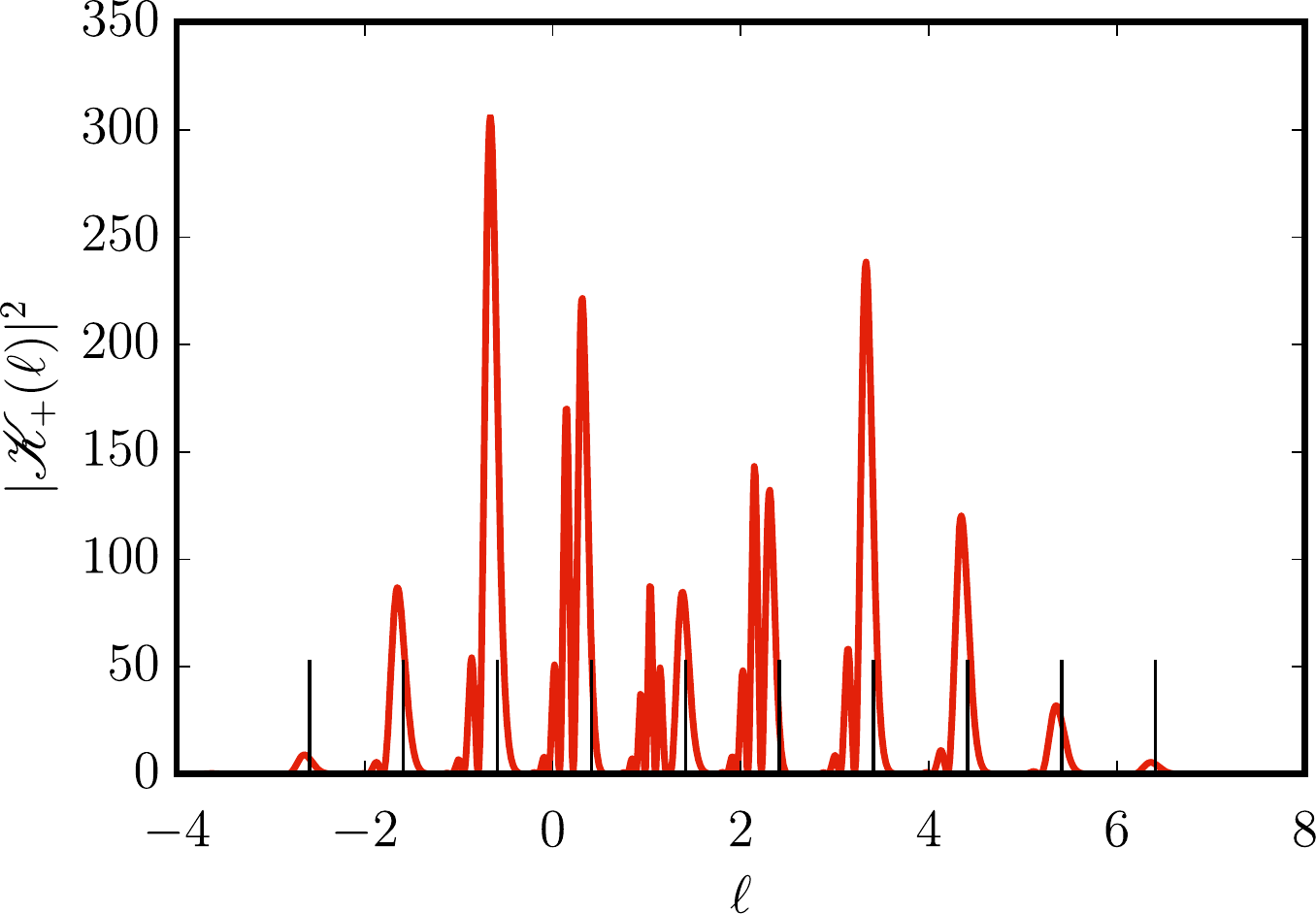}}
\includegraphics[width=0.49\columnwidth]{{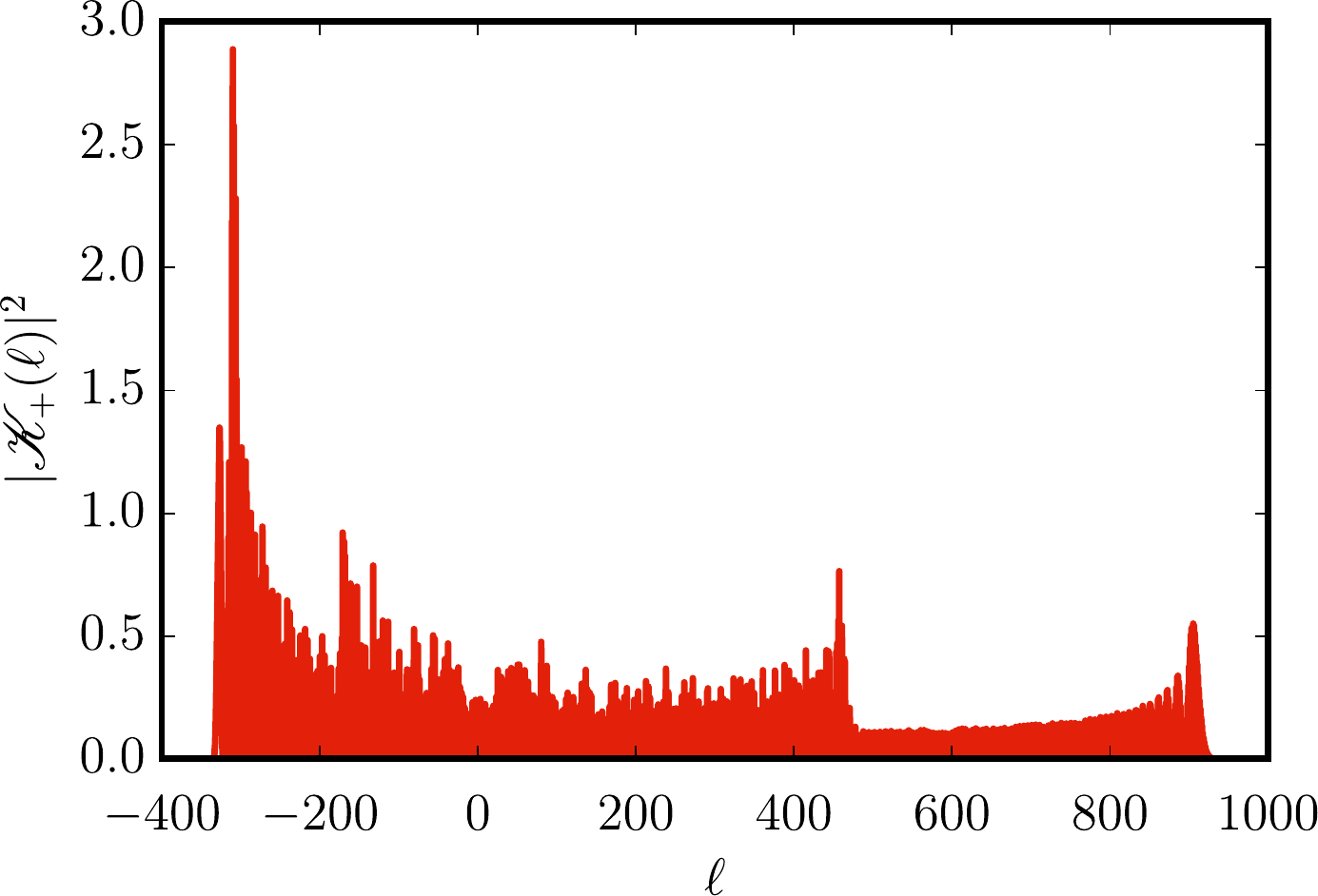}}
\caption{Spectral components of the Volkov state with a small (large) range of spectral components $\ell$
contributing in the left (right) panels.
In the left panel the black vertical lines indicate the positions of the
Zel'dovich levels at $\ell = n+\frac{ma_0^2}{8\omega\gamma}$ in the case of an infinite monochromatic plane wave.
One can make a clear connection between these Zel'dovich levels and the peaks in the Volkov spectrum
for a pulsed field.
In the right panel a large number of spectral components contribute the width of each level is larger than their separation which 
makes a clear identification of individual Zel'dovich levels difficult.
\label{fig:zeldovich}
}
\end{figure}

\begin{figure}[!t]
\centering
\includegraphics[width=0.49\columnwidth]{{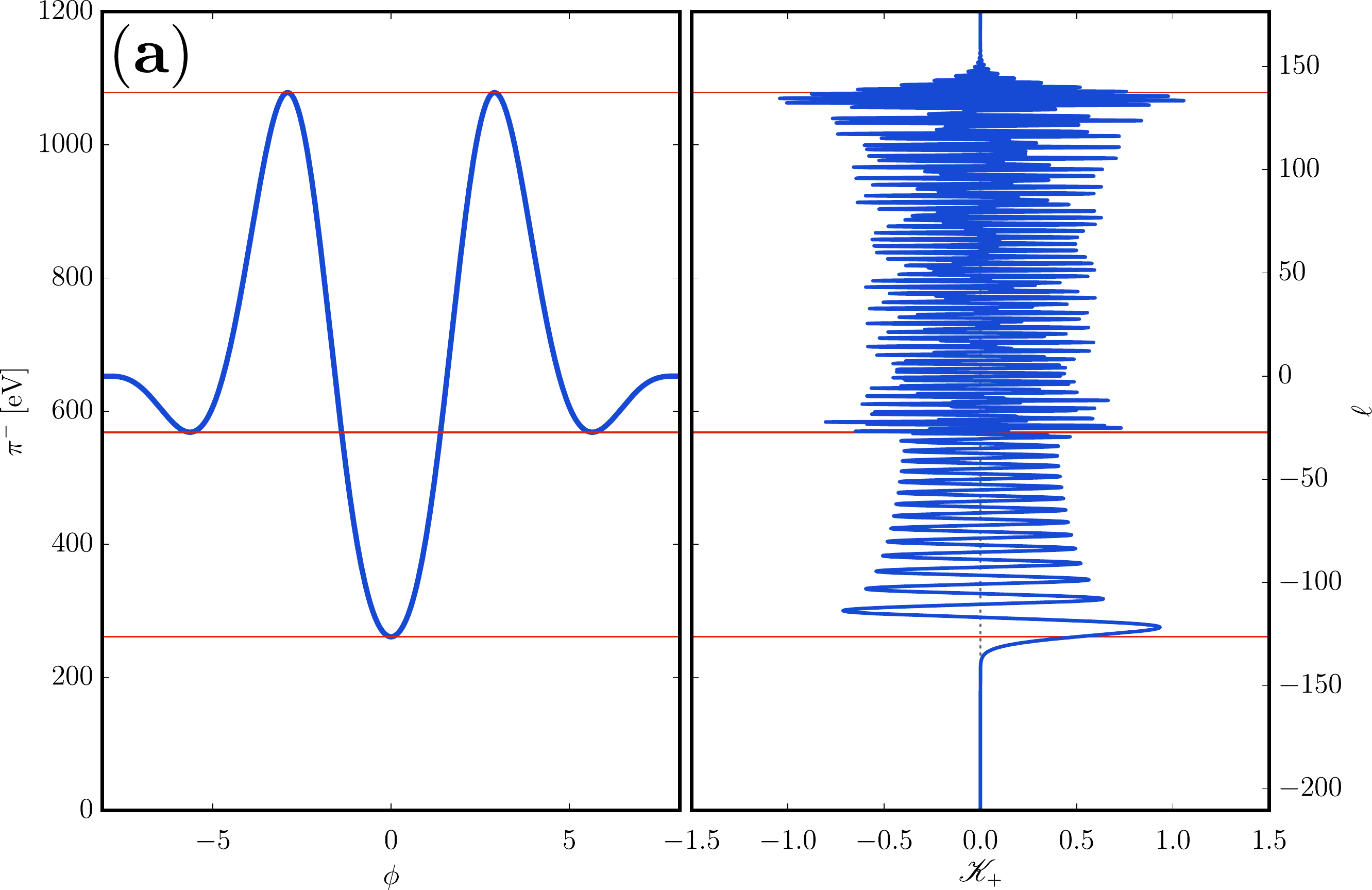}}
\includegraphics[width=0.49\columnwidth]{{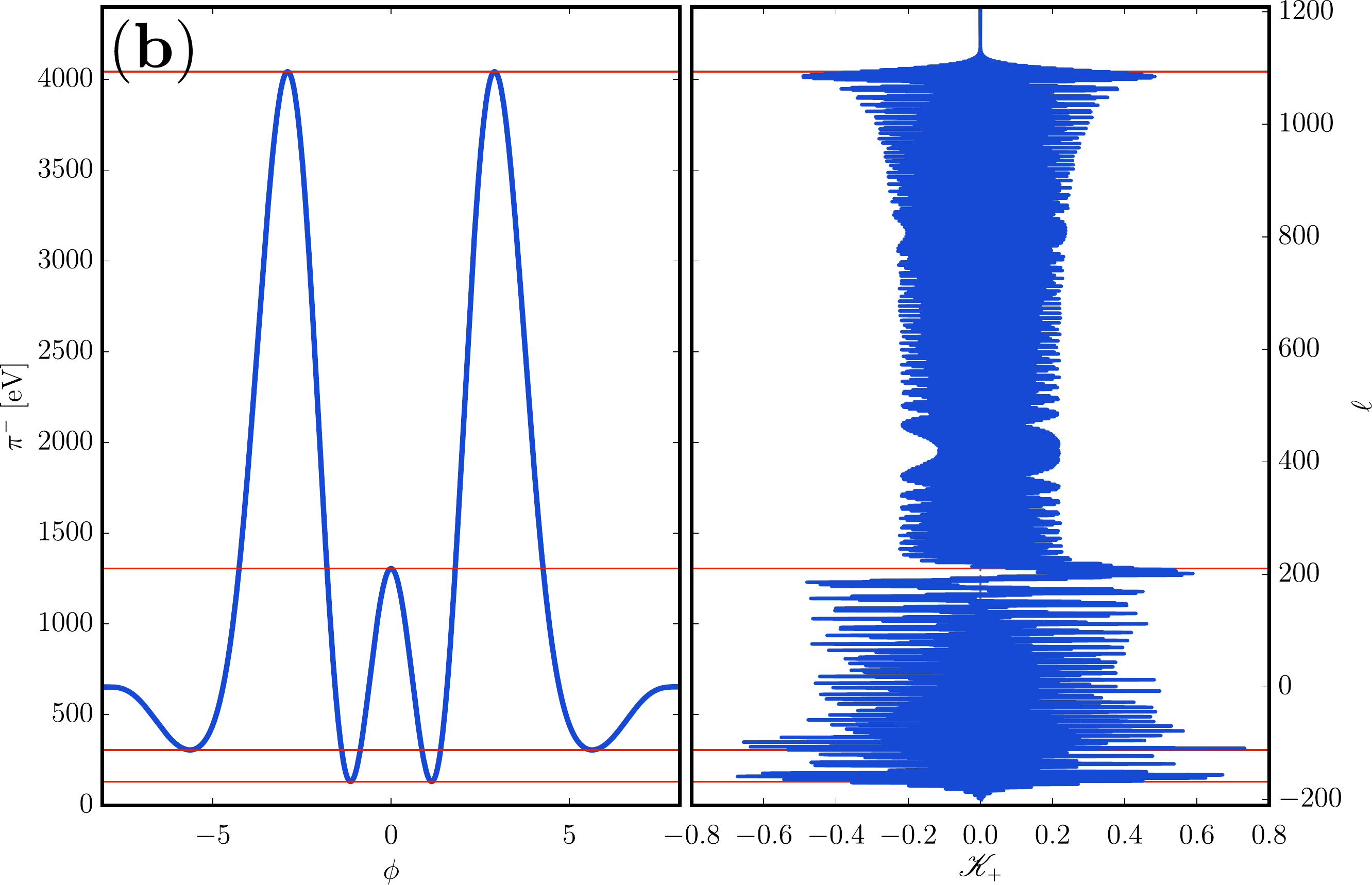}}

\includegraphics[width=0.49\columnwidth]{{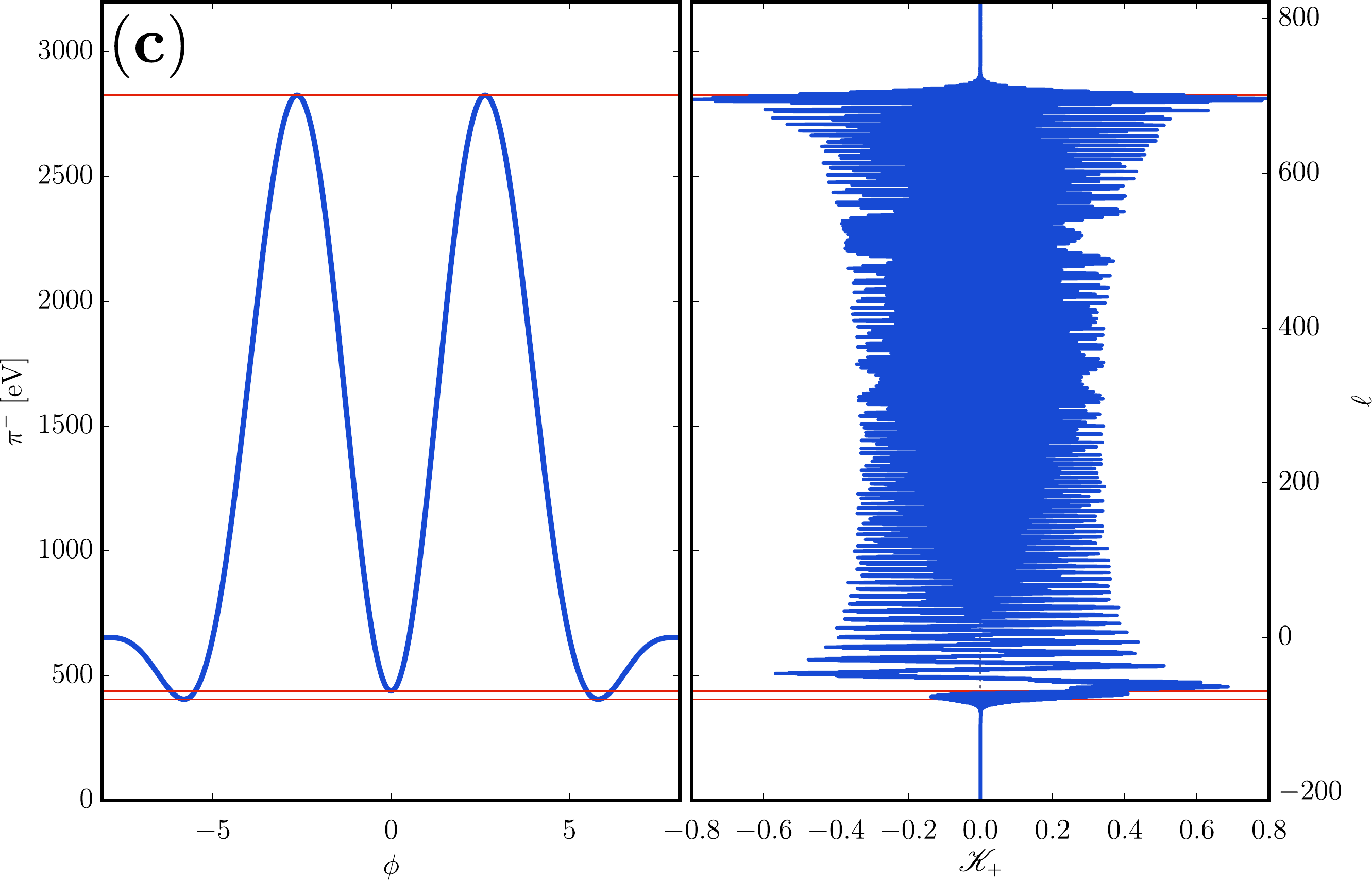}}
\includegraphics[width=0.49\columnwidth]{{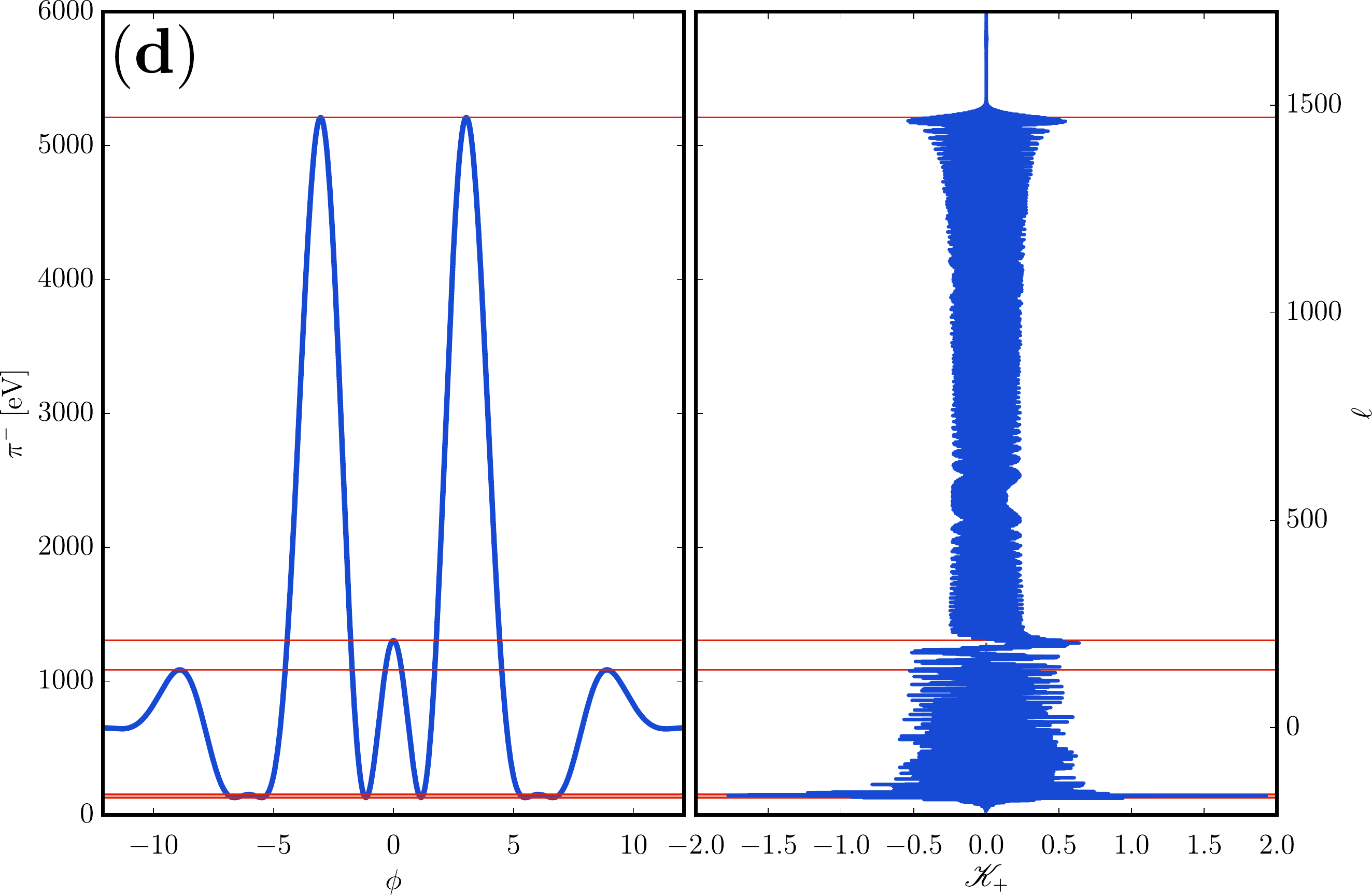}}
\caption{The light-front momentum component $\pi^-$ along the classical trajectory of an electron in a short laser pulse as a
function of the laser phase $\phi = \omega x^+$ (left panel) is compared to the spectral component
$\mathrsfs K_+(\ell)$ of the Volkov state in the same pulse (right panel). The electron counterpropagates the laser
pulse with initial energy of $1$ GeV at an angle of $\theta = 2/\gamma$.
Other parameters are as follows: 
(a) $a_0 = 1 $, $\plp = 8$, linear polarization;
(b) $a_0 = 5$, $\plp=8$, linear polarization;
(c) $a_0=5$, $\plp = 8$, circular polarization;
(d) $a_0=5$, $\plp = 12$, linear polarization.
}
\label{fig:trajectory:volkov}
\end{figure}

If one now starts from the well known case of infinite monochromatic plane waves (with the discrete level structure) and looks 
at modifications due to \textit{finite} pulse duration one finds a broadening of the Zel'dovich levels.
There are two different contributions to the level broadening: (i) a bandwidth broadening $\plp^{-1}$ due to the finite laser bandwidth
(ii) a non-linear, intensity dependent broadening $\sim \frac{m a_0^2}{\omega\gamma}$ due to the gradual change in the laser intensity in a pulsed field.
This second mechanism is related to the ponderomotive, i.e.~slowly varying, part of the non-linear phase $f_p$,
and can be seen as a gradual build-up of an intensity-dependent quasi-momentum as the laser intensity increases \cite{Narozhnyi:JETP1996}.
Depending on the bandwidth and the peak intensity either of the two mechanisms can be dominating the broadening of the Zel'dovich levels. The broadened Zel'dovich levels are shown in Fig.~\ref{fig:zeldovich}.

The scalar spectral components $\mathrsfs K_j(\ell)$ of a Volkov state also have a close relation
to the classical electron motion in a laser pulse.
The effective range of values of $\ell$ is determined by the classical dynamics \cite{Loetstedt:PRE2009}.
To be specific, the maximum and minimum values of the minus component of the kinetic four-momentum
determine the cut-off values of $\ell$ via $\ell_\mathrm{max/min} = (\pi^-_\mathrm{max/min} - p^-)/k^-$.
Beyond those cut-off values the functions $\mathrsfs K_j(\ell)$ drop to zero
exponentially fast. The reason is that only for $\ell_\mathrm{min} < \ell < \ell_\mathrm{max}$ the phase integrals in \eqref{eq.def.Kn}
posses real stationary phase points.
Also the local minima and maxima of $\pi^-$, which
appear during the course of the laser pulse, lead to pronounced structures in the Volkov state spectral component
$\mathrsfs K_+(\ell)$. This is caused by a sudden change in the number of stationary phase points related to fold-type caustics.
All the local extremal points of $\pi^-$ are indicated by red lines in Figure~\ref{fig:volkov:state:spatial}, connecting them to 
pronounced structures in the spectral components
$\mathrsfs K_j(\ell)$.
One could possibly observe those spectral features of the Volkov states by using laser-assisted Compton scattering of X-rays
\cite{Seipt:NJP2016}.}

\subsection{The Dirac-Volkov Propagator}

The laser-dressed Green's function (Dirac-Volkov propagator) is the solution of an inhomogeneous Dirac equation in the background
field with delta inhomogeneity, $[ i \slashed \partial - \slashed A -m  ]\: \mathcal G(x,y) = \delta(x -y)$.
Many different representations of this Dirac-Volkov propagator $\mathcal G(x,y)$ are known,
see e.g.~\cite{Eberly:PR1966,Brown:PR1964,Kibble:NPB1975,Baier:JETP1975b,Baier:JETP1976b,Mitter:ActaPhysAustr1975,Ritus:JSLR1985}.
A particularly useful representation is given in terms of the Ritus matrix functions $E_p$ as \cite{Ritus:JSLR1985}
\begin{align}
\mathcal G(x,y) &= \int \!  \frac{d^4p }{(2\pi)^4} \, E_p(x) G_0( p ) \bar E_p(y)\,,
\label{def.propagator}
\end{align}
where the free propagator in momentum space is given by $G_0(p) = (\slashed p + m) / (p^2-m^2 +i0^+)$.
The Dirac-Volkov propagator is needed for the calculation of higher-order processes in laser fields
such as nonlinear Double Compton scattering \cite{Lotstedt:PRL2009,Seipt:PRD2012,Mackenroth:PRL2013},
laser assisted Bremsstrahlung \cite{Loetstedt:PRL2007}
or laser assisted pair production in the field of a nucleus \cite{Loetstedt:PRL2008},
as well as the calculation of radiative corrections such as the electron mass operator \cite{Ritus:AnnPhys1972}.
It is worth noting that the background field determines the pole structure of the propagator, with an infinite series of poles for infinite monochromatic plane waves and just a single pole plus finite resonances for short pulses \cite{Seipt:PRD2012,diss:Seipt,Ilderton:PRD2013,Lavelle:2013}.

\subsection{Strong-Field Feynman Rules}

In order to calculate the S matrix elements of high-intensity QED processes one may employ the Feynman diagram technique
in coordinate space.
The Feynman rules for high-intensity {QED} in the Furry picture
can be summarized as follows \cite{Mitter:ActaPhysAustr1975}:
\begin{enumerate}
\item External incoming or outgoing electrons with momentum $p$
are represented by laser dressed Volkov wave function $\Psi_p(x)$ or $\bar \Psi_p(x)$, respectively.
For incoming and outgoing positrons one uses the corresponding negative energy Volkov wave function
$\bar \Psi_{-p}(x)$ and $\Psi_{-p}(x)$.
\item An internal fermion line corresponds to the Dirac-Volkov propagator $\mathcal G(x,y)$.
\item Internal and external photon lines are translated into
the free photon propagator and the free photon wave functions, respectively, see e.g.~Ref.~\cite{book:Peskin}.
\item Each interaction vertex corresponds to a factor $-i e\gamma^\mu$ and an integral over $d^4x$.
\item Symmetry factors for identical particles etc.~are the same as in usual {QED}.
\end{enumerate}

\begin{figure}
$$ \qquad \bf (a) \hspace*{0.4\columnwidth} (b) \hspace*{0.3\columnwidth} $$
\begin{fmffile}{./graphs}
\begin{align*}
	\parbox{40mm}{
	\begin{fmfgraph*}(70,24)
	 \fmfleft{l,d}
	 \fmfright{r,d}
	 \fmftop{d,d,t1,t2,d}
	 \fmf{dbl_plain,tension=1.25}{l,v1}
	 \fmf{dbl_plain,label.side=right}{v1,r}
	 \fmffreeze
	 \fmf{photon}{v1,t2}
	 \fmfdot{v1}
	 \fmflabel{$k'$}{t2}
	 \fmflabel{$p$}{l}
	 \fmflabel{$p'$}{r}
	 \end{fmfgraph*}}
&\qquad& 
	\parbox{30mm}{
	\begin{fmfgraph*}(70,24)
	\fmfstraight
	 \fmfleft{l,d}
	 \fmfright{r,d}
	 \fmf{dbl_plain,label.side=right}{l,v1,v2,r}
	 \fmffreeze
	 \fmftop{d,d,t1,t2}
	 \fmf{photon}{v1,t1}
	 \fmf{photon}{v2,t2}
	 \fmfdot{v1,v2}
	 \fmflabel{$k_1$}{t1}
	 \fmflabel{$k_2$}{t2}
	 \fmflabel{$p$}{l}
	 \fmflabel{$p'$}{r}
	 \end{fmfgraph*}}
	\qquad  + \qquad 
	\parbox{30mm}{
	\begin{fmfgraph*}(70,24)
	\fmfstraight
	 \fmfleft{l,d}
	 \fmfright{r,d}
	 \fmf{dbl_plain,label.side=right}{l,v1,v2,r}
	 \fmffreeze
	 \fmftop{d,d,t1,t2}
	 \fmf{photon}{v1,t1}
	 \fmf{photon}{v2,t2}
	 \fmfdot{v1,v2}
	 \fmflabel{$k_2$}{t1}
	 \fmflabel{$k_1$}{t2}
	 \fmflabel{$p$}{l}
	 \fmflabel{$p'$}{r}
	 \end{fmfgraph*}} 
\end{align*}
\end{fmffile}
\caption{Strong-field Feynman diagrams in the Furry picture for Non-linear Compton scattering (a) and Non-linear Double Compton scattering (b). Double lines represent laser dressed electron wave functions and propagators for external and internal lines, respectively.}
\label{fig:diagrams}
\end{figure}
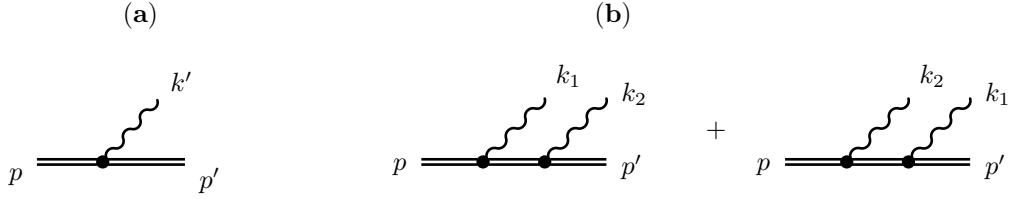

Let us look at few examples.
Firstly, the S matrix for non-linear Compton scattering, i.e.~the emission of a high-energy photon by an electron moving in
an intense laser field, given  in Fig.~\ref{fig:diagrams} (a).
This diagram is are forbidden in ordinary QED without the background field
because it is impossible to fulfil four-momentum conservation.
The background field provides the missing four-momentum to make this process possible.
Diagram (a) translates into the formula
\begin{align} \label{eq:SNLC}
S =  \int \! d^4 x \: \bar \Psi_{p'} (x) [-ie \gamma^\mu \varepsilon^*_\mu e^{ik'\cdot x}] \Psi_{p}(x) \,,
\end{align}
where the wave function of the emitted photon with momentum $k'$ and polarization vector $\varepsilon'$ is given by 
$\varepsilon^*_\mu e^{ik'\cdot x}$. The non-linear Compton scattering is discussed in some more detail in Section \ref{sect:compton}.

The second example is the non-linear Double Compton scattering in (b), where two photons are emitted coherently by the same electron.
Because the two emitted photons are indistinguishable the S matrix is represented by two diagrams with the outgoing photon
lines interchanged.
The S matrix for this process reads
\begin{align*}
S & = - e^2 \int \! d^4x d^4y \: \bar \Psi_{p'}(x) \, \slashed \epsilon^*_2 e^{i k_2\cdot y} \,
									 \mathcal G(y,x|A) \, \slashed \epsilon^*_1 e^{i k_1\cdot x} \, \Psi_p(x)	\\
			& \quad -e^2 \int \! d^4x d^4y \: \bar \Psi_{p'}(x) \, \slashed \epsilon^*_1 e^{i k_1\cdot y} \,
									 \mathcal G(y,x|A) \, \slashed \epsilon^*_2 e^{i k_2\cdot x} \, \Psi_p(x) \,.
\end{align*}
The pole structure of the Dirac-Volkov propagator strongly affects the spectrum of the emitted photons
\cite{Oleinik:JETP1967,Lotstedt:PRL2009,Seipt:PRD2012}.

\section{Non-linear Compton Scattering}

\label{sect:compton}

The process of non-linear Compton scattering in high-intensity lasers has been investigated theoretically since the early 1960's
\cite{Nikishov:JETP1964a,Nikishov:JETP1964b,Nikishov:JETP1965,Narozhnyi:JETP1965}, but in most studies
infinite monochromatic plane wave laser fields or constant crossed fields have been assumed.
But as mentioned above, modern laser systems produce very short high-intensity pulses with only a few cycles oscillations of the
laser's electric field.
The influence of the short laser pulse duration on the non-linear Compton process has been investigated in the recent years,
see for instance Refs.~\cite{Narozhnyi:JETP1996,Boca:PRA2009,Seipt:PRA2011,Mackenroth:PRA2011,Krajewska:PRA2012,Seipt:PRA2015,Kharin:PRA2016,Seipt:JPP2016}.
We shall here discuss some of the short pulse effects on the non-linear Compton spectra.
{Similar effects can be seen also for other processes occuring in strong and short laser pulses, for instance
in the non-linear Breit-Wheeler pair production \cite{Nousch:PLB2012,Krajewska:PRA2012b,Titov:PRL2012,Titov:PRD2016,Meuren:PRD2016,Jansen:PRD2016}.}

\begin{figure}[!b]
\centering
\begin{minipage}{0.49\columnwidth}
\includegraphics[width=\textwidth]{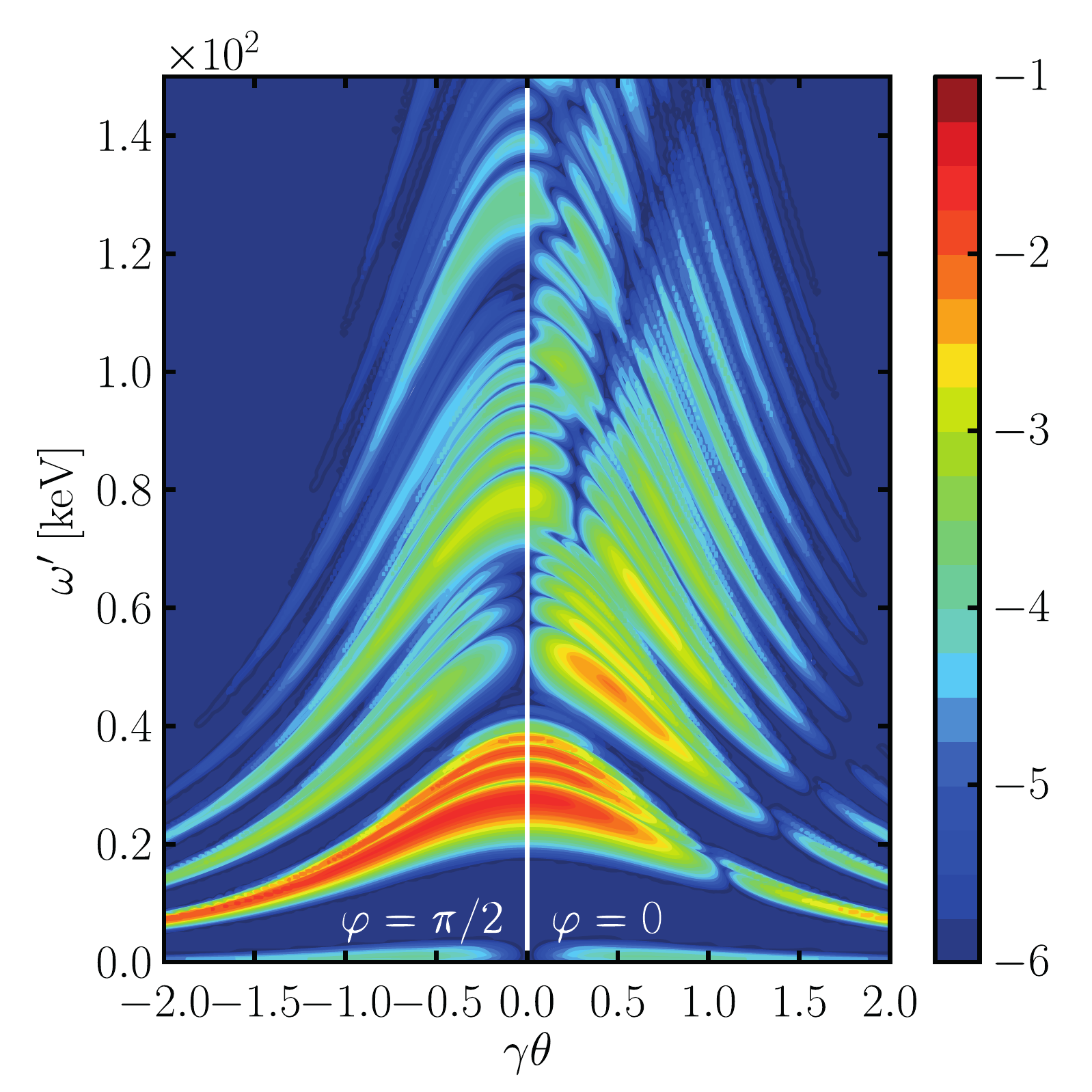}
\end{minipage}
\begin{minipage}{0.35\columnwidth}
\includegraphics[width=\textwidth]{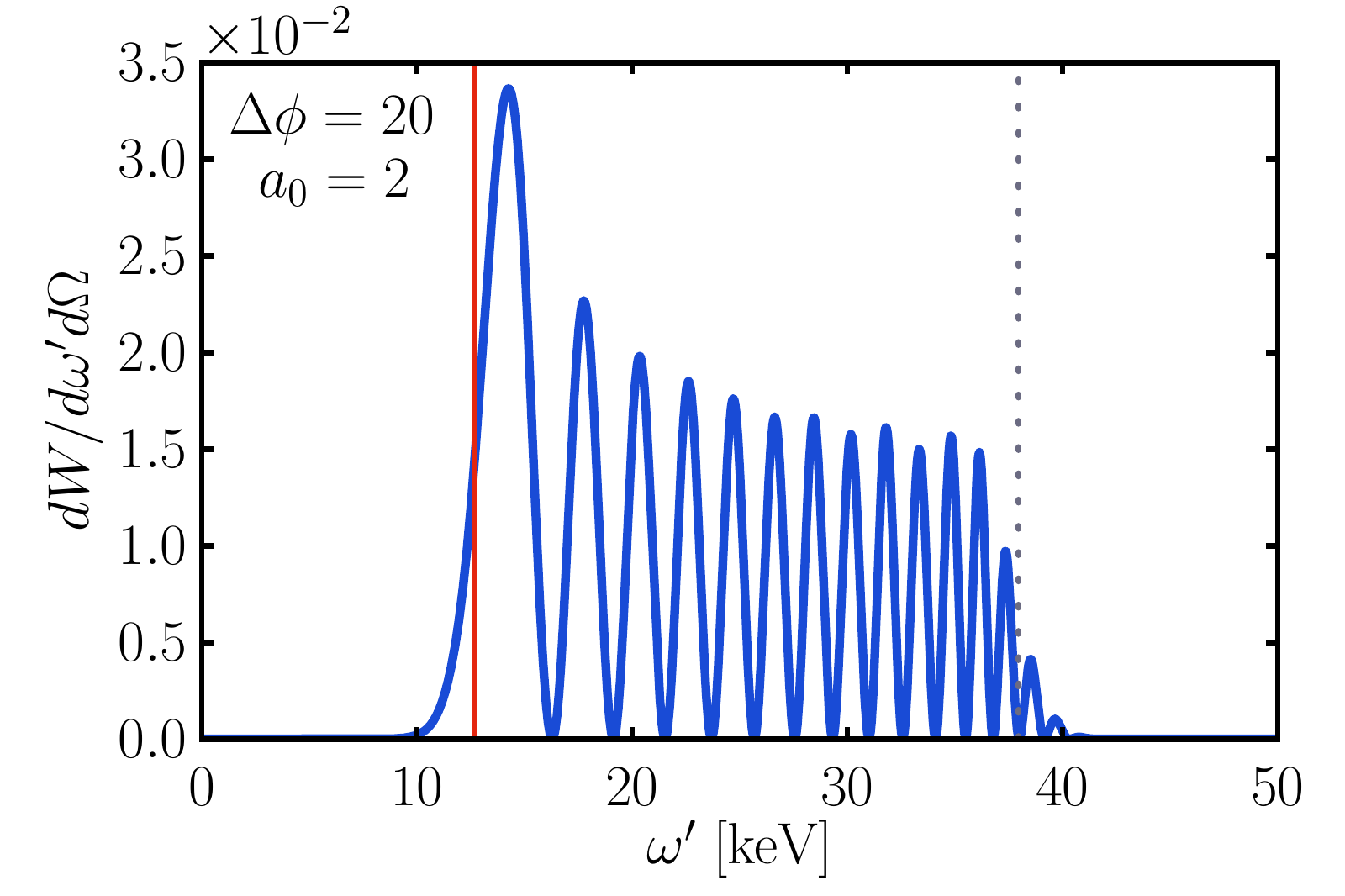}
\includegraphics[width=\textwidth]{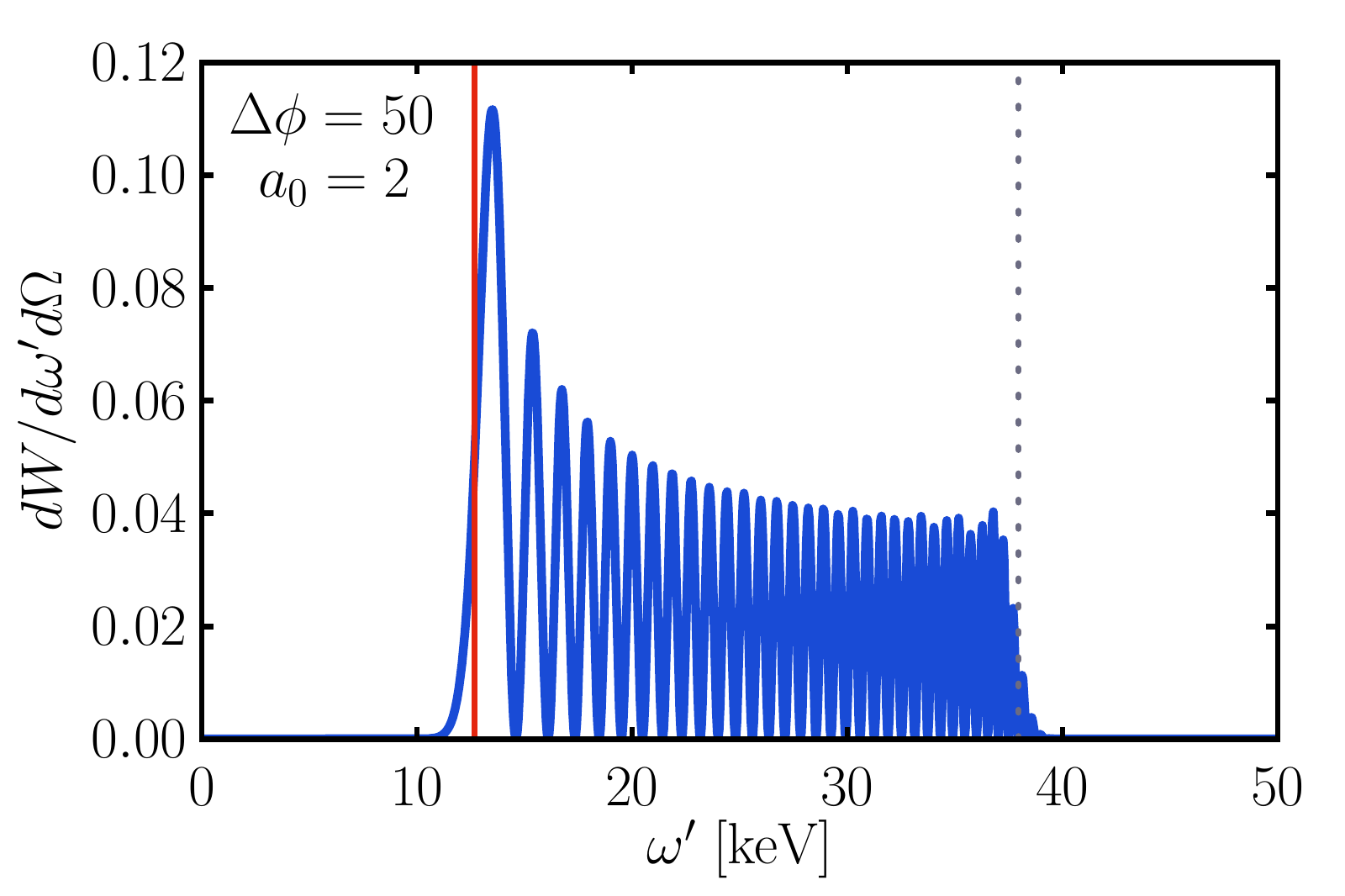}
\end{minipage}
\caption{Spectra for non-linear Compton scattering in a short laser pulse. Left panel: angular and energy differential spectra for linearly polarized laser light in the plane of the laser polarization  ($\varphi=0$) and perpendicular to it ($\varphi = \pi/2$).
Right panels: Line-outs of the on-axis spectrum for circularly polarized laser light show the broadening of the spectral lines in short pulses.
In the case of an infinitely long laser wave the spectrum would be located just at the red vertical line.
}
\label{fig:nlc1}
\end{figure}

The differential photon emission probability is given by
$ dW = | S |^2 \, d\Pi / 2p^+ $,
with the S matrix element from \eqref{eq:SNLC}, and
where the Lorentz invariant phase space of the final particles is conveniently parametrized by
$ 
d\Pi = \frac{d p'^+ d^2\mathbf p^\perp}{2(2\pi)^3 p'^+} \frac{\omega' d \omega' d\Omega }{2(2\pi)^3 }  \,,
$
yielding
\begin{align} \label{eq:probability}
\frac{dW}{d\omega' d\Omega} = \frac{\alpha \, \omega' \:  \langle |\mathcal M|^2\rangle  }{16\pi^2 \, (kp)(kp')} \,, & & \mathcal M =  \int d \phi \: \bar u_{p'} \Gamma(\phi) u_p \: e^{i\int d\phi \frac{k'\cdot \pi}{k\cdot p'}}
\end{align}
where $\langle...\rangle$ refers to the average and summation over all initial and final particle polarization states, respectively.
In the exponent of the amplitude
$\mathcal M$ we rediscover the classical kinetic momentum $\pi$, which is now
projected onto the four-momentum of the emitted photon.

When the probability \eqref{eq:probability} is analysed for an infinite monochromatic plane wave one finds
a discrete spectrum of emitted harmonics, which can be understood in terms of \textit{quasi-momentum}
conservation $q+nk = q'+k'$. The emitted photons are dominantly emitted into a narrow cone
with opening angle $\theta \sim \gamma^{-1}$ for $a_0\lesssim1$ and $\sim a_0 \gamma^{-1}$ for $a_0 > 1$,
and with harmonic frequencies
$\omega_n' \simeq 4n\gamma^2\omega / (1 + \gamma^2\theta^2 + 2n b_0+a_0^2/2)$.
For $a_0 \gg1$ on the order of $n\sim a_0^3$ harmonics contribute to the spectrum.
In non-linear Compton scattering with a \textit{short laser pulse}
one observes a broadening of the spectral lines (see Fig.~\ref{fig:nlc1}),
similar to the broadening to the Zel'dovich level structure of the Volkov states discussed in Section \ref{sect:volkov}.
First of all, in a short laser pulse the harmonics acquire a bandwidth
$\Delta \omega' / \omega' \sim \Delta \omega / \omega \sim \plp^{-1}$ due to the bandwidth of the laser.
But there is also a intensity dependent ponderomotive broadening effect related to the gradual build-up of quasi-momentum
as the electron enters the laser pulse. The ponderomotively broadened spectral lines have a (relative) bandwidth
 $\sim (a_0^2/2)/(1 + a_0^2/2)$.
One sees in Fig.~\ref{fig:nlc1} that the broadened spectral lines consist of a large number of $\mathcal O(a_0^2 \, \plp)$
subsidiary peaks. This can be explained as the interference of radiation emitted at the leading and trailing slopes of the pulse envelope
\cite{Kharin:PRA2016}.
It turns out that this ponderomotive broadening effect always occurs whenever the dephasing parameter
$a_0^2 \, \plp\gtrsim 1$.
This is true also for quite weak laser pulses $a_0\ll1$ as long as the spectral bandwidth of the laser
$\plp^{-1}$ is sufficiently small \cite{Hartemann:PRL2010}.

\begin{figure}[!t]
\centering
\includegraphics[width=0.6\columnwidth]{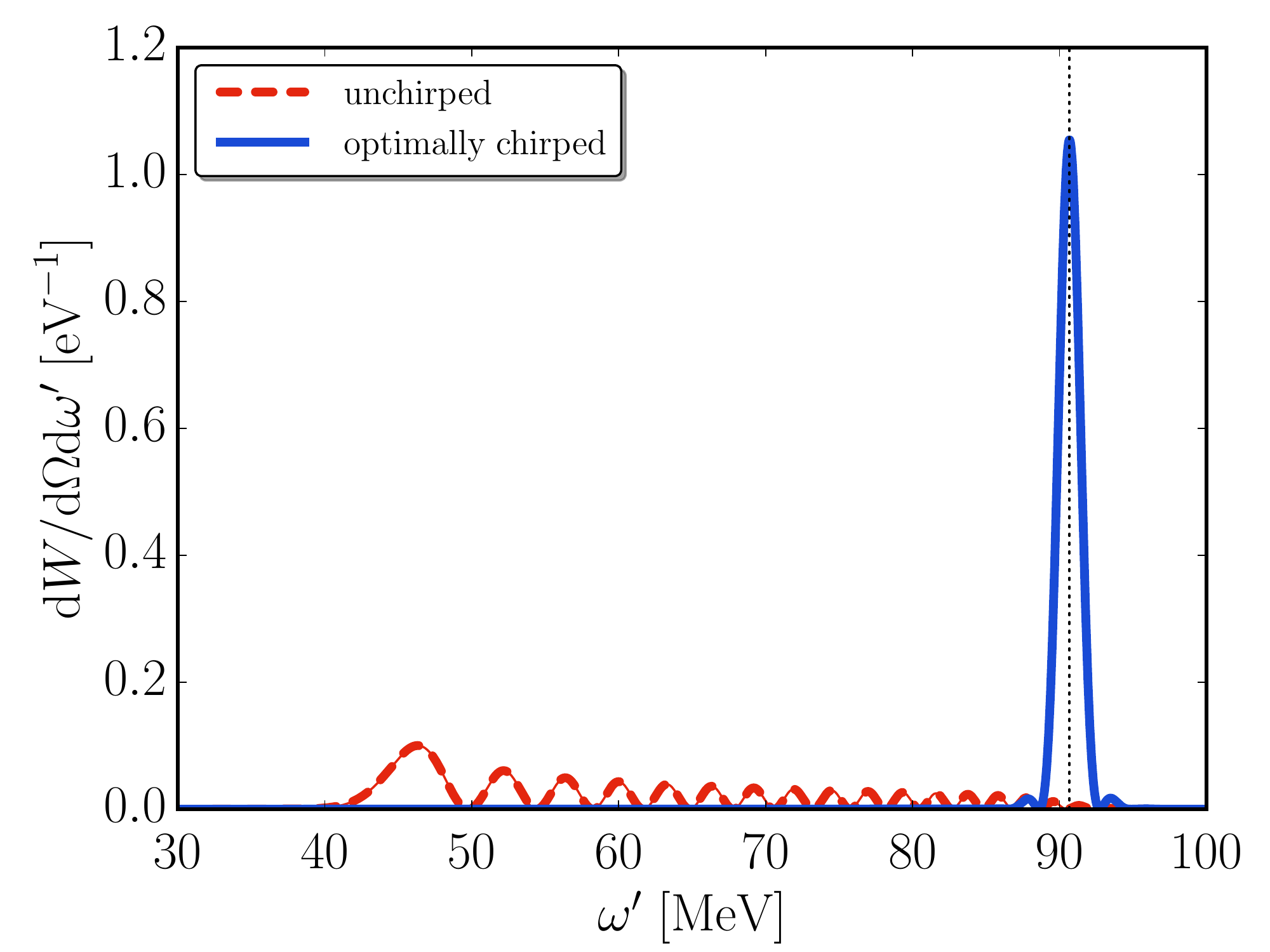}
\caption{By using a {chirped} laser pulse with, the spectral broadening of the backscattered light (red curve) can be compensated (blue curve) allowing to operate inverse Compton sources at high laser intensity ($a_0=1.5$) for increased brightness.}
\label{fig:compensation}
\end{figure}

{This broadening effect, while being an interesting phenomenon in the interplay between bandwidth and intensity effects
will be a severe limitation for the spectral brightness of inverse Compton backscattering x- and gamma-ray sources.
One way of dealing with the broadening is to limit the laser intensity to an acceptable level of broadening for a specific application \cite{Hartemann:PRL2013},
which limits the brightness of the source, of course.}
A more advanced possibility is to use \textit{chirped} laser pulses, i.e. with a time-dependent frequency.
It has been shown theoretically that with properly chirped laser pulses (which are spectrally blue-shifted at the pulse centre where the intensity is highest) the spectral broadening can be compensated
\cite{Ghebregziabher:PRSTAB2013,Terzic:PRL2014,Seipt:PRA2015,Rykovanov:PRSTAB2016}, see Fig.~\ref{fig:compensation}.
This opens an avenue towards bright narrowband
hard-gamma radiation sources operated at large laser intensities $a_0 \gtrsim1$.

\section{Summary}
In this article an overview of the theory of high-intensity QED was given with focus on effects due to the short pulse duration.
Both the solutions of the classical equations of motion, as well as the Volkov wave functions and the Dirac-Volkov propagator in
the case of high-field QED were presented. The spectrum of non-linear Compton scattering in
short intense pulses was discussed.
The intensity dependent spectral broadening can be compensated by
chirping the laser pulse.

\section*{Appendix A: Light-Front Coordinates}
\label{app:lightfront}

\begin{figure}
\begin{center}
\includegraphics[width=0.5\textwidth]{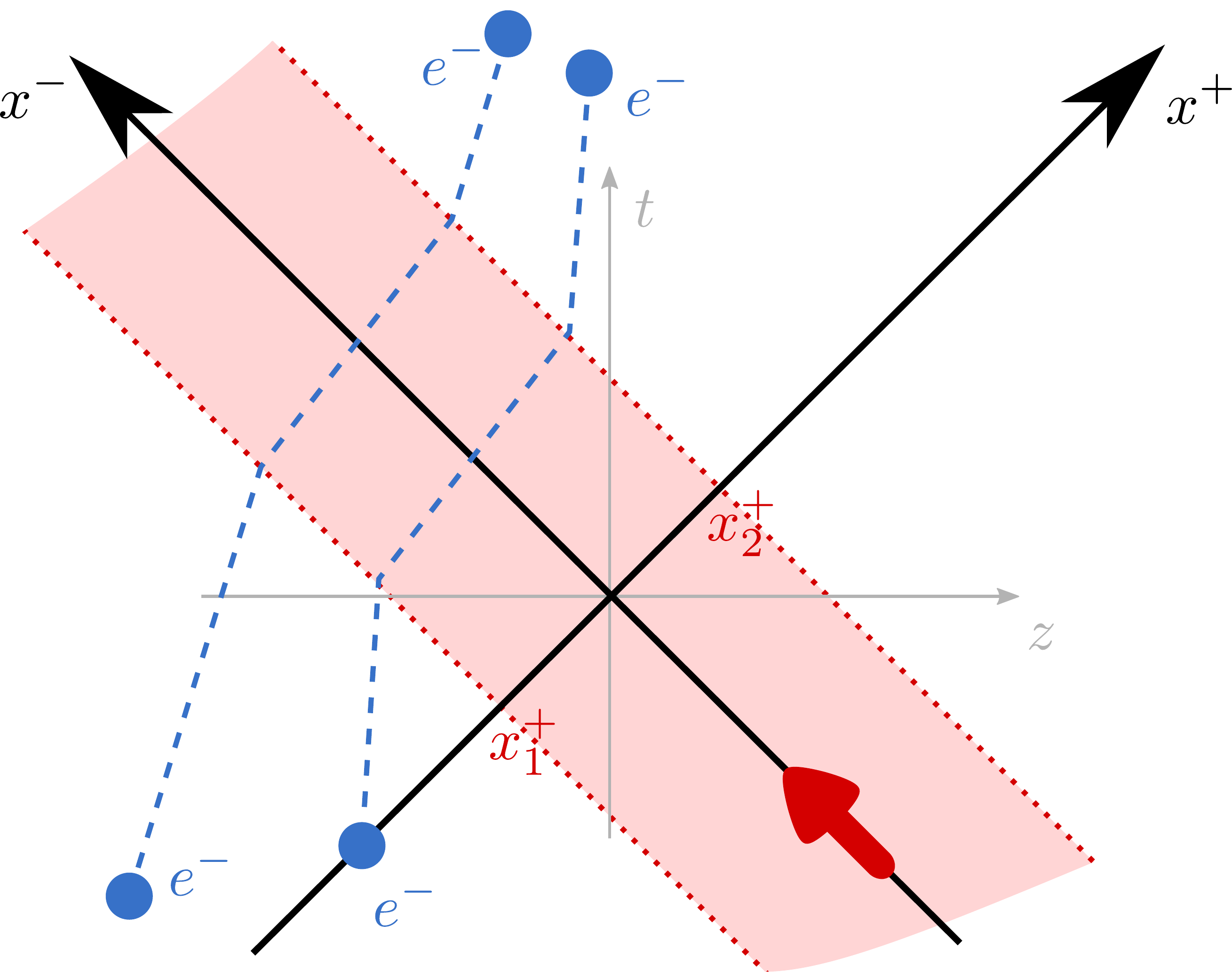}
\end{center}
\caption{The laser pulse (red) propagates along the negative $z$ axis. All electrons enter (leave) the laser at the same value of the light-front
variable $x^+= x^+_1$ ($x^+_2$).}
\label{fig:lightfront}
\end{figure}

Because the laser's wave-vector (four-momentum vector) $k^\mu$ is a light like vector with $k^2=0$
it is very convenient to use so-called light-front co-ordinates to describe the dynamics of charged particles in
laser background fields. Let us assume that the laser pulse propagates along the negative $z$-axis, $k=(\omega,0,0,-\omega)$,
such that it's phase argument $k\cdot x = \omega (t+z) = \omega x^+$. That means the background field depends only on
the single light-front variable $x^+ = t+z$. All particles will enter (leave) the laser pulse at the same value of the the light-front time coordinate $x^+$, see Fig.~\ref{fig:lightfront}.
The light-front components of a generic four-vector with Cartesian coordinates
$(  b^{\mu} ) = (b^0,b^1,b^2,b^3)$ are defined as
\begin{align}
b^{-}=b^0-b^3\,,\qquad b^{+}=b^0+b^3\,,\qquad\mathbf{b}^{\perp}=(b^1,b^2)\,.
\end{align}
The inverse coordinate transformation is given by $b^{0}=\frac{1}{2}(b^{+}+b^{-})$ and
$b^{3}=\frac{1}{2}(b^{+}-b^{-})$.
We immediately find that the laser four-momentum has only a single non-vanishing component in light-front coordinates: $k^-=2\omega$
The metric tensor in light-front coordinates, using the
arrangement of components $(b^{\mu})  =  (b^{+},b^{-},\mathbf{b^{\perp}})$,
has non-diagonal components
\begin{align}
( g_{\mu\nu}) & {\displaystyle \ =\ }  \left(\begin{tabular}{cccc}
 \mbox{\ 0\ }  &  \mbox{\ \ensuremath{\frac{1}{2}}\ }  &  \mbox{\ 0\ }  &  \mbox{\ 0\ } \\
\ensuremath{\frac{1}{2}} &  0  &  0  &  0 \\
0  &  0  &  $-1$  &  0 \\
0  &  0  &  0  &  $-1$ \end{tabular}\right)
\,, \quad & &
( g^{\mu\nu} )   {\displaystyle \ =\ }  \left(\begin{tabular}{cccc}
 \mbox{\ 0\ }  &  2  &  \mbox{\ 0\ }  &  \mbox{\ 0\ } \\
2&  \mbox{\ 0\ }  &  0  &  0 \\
0  &  0  &  $-1$  &  0 \\
0  &  0  &  0  &  $-1$ \end{tabular}\right)
\label{eq.def.metric} \,.
\end{align}
The covariant components of a four-vector are related to the contravariant components as
\begin{align*}
b_- = \frac{1}{2}\, b^+ \,, \qquad b_+ = \frac12 \, b^-\,, \qquad \mathbf b_\perp = - \mathbf b^\perp\,,
\end{align*}
and scalar products between two four-vectors read
\begin{align*}
x \cdot y &= x^+ y_+ + x^-y_- + \mathbf x^\perp \cdot \mathbf y_\perp 
		= \frac{1}{2} x^+ y^- + \frac{1}{2} x^- y^+ -  \mathbf x_\perp \cdot \mathbf y_\perp \,.
\end{align*}
The Lorentz invariant integration measure in light-front coordinates is given by
\begin{align*}
\sqrt{-g} \, d^{4}x = \frac{1}{2} \, d x^{+} d x^{-} d^{2}\mathbf{x}_{\perp}\,,
\end{align*}
with $\sqrt{-g}= \sqrt{-{\rm det} \, g_{\mu\nu} }=1/2$ as the determinant of the metric tensor.
For a four-dimensional delta distribution one writes in light-front form
\begin{align*}
\delta^4(p) = \frac{1}{\sqrt{-g}} \delta(p^+) \delta(p^-) \delta(p^1) \delta(p^2) 
= 2 \, \delta^3(\mathsf{p}) \delta(p^-)\,,
\end{align*}
where we introduced the short-hand notation $\mathsf{p} \equiv (p^+,\mathbf p^\perp)$.
%
The free particle dispersion relations (mass shell conditions) read in light-front coordinates
\begin{align*}
p^- =\frac{ \mathbf{p}_{\perp}^{2} + m^{2}}{p^+ } \,, \qquad
k^- = \frac{\mathbf k_\perp^2}{k^+} 
\end{align*}
for massive and massless particles, respectively.
The Lorentz invariant on-shell phase space element is
\begin{align*} \label{eq:light-front:phasespace}
\int\!  \frac{d^4p}{(2\pi)^4}\, (2\pi) \delta(p^2-m^2) &= \frac{d^3\sf p}{(2\pi)^3 2 p^+}\,,
\end{align*}
with $d^3\mathsf{p} \equiv dp^+ d p^1 d p^2$.
Thus the Lorentz invariant on-shell delta distributions are given by $(2\pi)^3 2p^+ \delta^3(\mathsf{p} -\mathsf{p}') $.

The light-front components of the Dirac matrices $\gamma^\mu$   obey the anti-commutation relations
$\{\gamma^\mu,\gamma^\nu\} = 2 g^{\mu\nu}$ with the metric tensor \eqref{eq.def.metric}. In particular
it follows that $(\gamma^+)^2 = (\gamma^-)^2 = 0$.

\begin{footnotesize}



\end{footnotesize}
\end{document}